\documentclass[journal,10pt]{IEEEtran}
\usepackage{booktabs}
\usepackage{balance}
\usepackage{xcolor}
\usepackage{soul}
\usepackage{array}
\newcolumntype{M}[1]{>{\centering\arraybackslash}m{#1}}
\usepackage{epsfig}
\usepackage{color}
\usepackage{amsmath}
\makeatletter
\renewcommand*\env@matrix[1][*\c@MaxMatrixCols c]{%
  \hskip -\arraycolsep
  \let\@ifnextchar\new@ifnextchar
  \array{#1}}
\makeatother

\usepackage{graphicx}
\usepackage{epstopdf}
\usepackage{algorithm}
\usepackage{algorithmic}
\usepackage{eucal}
\usepackage{enumerate}
\usepackage{cite}
\usepackage{amsfonts}
\usepackage{makecell}
\usepackage{lettrine}
\usepackage[belowskip=0pt,aboveskip=-15pt]{caption}
\usepackage{amssymb}
\usepackage{blindtext}
\usepackage{multirow}
\usepackage{lipsum}
\usepackage{amsbsy}
\usepackage[font=small]{caption}
\def\BibTeX{{\rm B\kern-.05em{\sc i\kern-.025em b}\kern-.08em T\kern-.1667em\lower.7ex\hbox{E}\kern-.125emX}}
\usepackage[font=small,skip=5pt]{caption}
\usepackage{verbatim}
\usepackage{amsmath}
\usepackage{amsfonts}
\usepackage{amsthm}
\usepackage{algorithm}
\usepackage{algorithmic}
\usepackage{caption}
\usepackage{subcaption}
\usepackage{float}
\usepackage{bbm}

\newcommand{\Ac}{\ensuremath{\mathcal{A}}}

\newcommand{\Rc}{\ensuremath{\mathcal{R}}}

\newcommand{\Rbb}{\ensuremath{\mathbb{R}}}

\newcommand{\Er}{\mathbf{E}}

\newcommand{\lb}{\ensuremath{\left(}}
\newcommand{\rb}{\ensuremath{\right)}}
\newcommand{\lbr}{\ensuremath{\left\{}}
\newcommand{\rbr}{\ensuremath{\right\}}}
\newcommand{\lsb}{\ensuremath{\left[}}
\newcommand{\rsb}{\ensuremath{\right]}}

\newcommand{\bcr}{\begin{color}{red}}
\newcommand{\bcb}{\begin{color}{blue}}
\newcommand{\ec}{\end{color}}

\newtheorem{theorem}{Theorem}

\usepackage{caption}
\usepackage{float}
\begin{document}
%\setlength{\pdfpagewidth}{8.5in}
%\setlength{\pdfpageheight}{11in}
%
% paper title
% can use linebreaks \\ within to get better formatting as desired
\title{Multi-Agent Task Assignment in Vehicular Edge Computing: A Regret-Matching Learning-Based Approach}
\author{\IEEEauthorblockN{Bach Long Nguyen, Duong D. Nguyen, Hung X. Nguyen, Duy T. Ngo, and Markus Wagner}\\

\thanks{Bach Long Nguyen, Hung X. Nguyen and Markus Wagner are with the School of Computer Science, The University of Adelaide, Adelaide, SA 5005, Australia (email: long.nguyen@adelaide.edu.au, hung.nguyen@adelaide.edu.au, markus.wagner@adelaide.edu.au).

Duong D. Nguyen is with the School of Electrical and Electronic Engineering, The University of Adelaide, Adelaide, SA 5005, Australia (email: duong.nguyen@adelaide.edu.au).

Duy T. Ngo is with the School of Electrical Engineering and Computing, The University of Newcastle, Callaghan, NSW 2308, Australia (email: duy.ngo@newcastle.edu.au).

}% <-this % stops a space

}
%\markboth{IEEE Transactions on Emerging Topics in Computational Intelligence,~Vol.~XX, No.~XX, XXX~2022}
{}

\maketitle

\begin{abstract}
Vehicular edge computing has recently been proposed to support computation-intensive applications in Intelligent Transportation Systems (ITS) such as self-driving cars and augmented reality. Despite  progress in this area, significant challenges remain to efficiently allocate limited computation resources to a range of time-critical ITS tasks. To this end, the current paper develops a new task assignment scheme for vehicles in a highway. Because of the high speed of vehicles and the limited communication range of road side units (RSUs), the computation tasks of participating vehicles are to be dynamically migrated across multiple servers. We formulate a binary nonlinear programming (BNLP) problem of assigning computation tasks from vehicles to  RSUs and a macrocell base station. To deal with the potentially large size of the formulated optimization problem, we develop a distributed multi-agent regret-matching learning algorithm. Based on the regret minimization principle, the proposed algorithm employs a forgetting method that allows the learning process to quickly adapt to and effectively handle the high mobility feature of vehicle networks. We theoretically prove that it converges to the correlated equilibrium  solutions of the considered BNLP problem. Simulation results with practical parameter settings show that the proposed algorithm offers the lowest total delay and cost of processing tasks, as well as utility fairness among agents. Importantly, our algorithm converges much faster than existing methods as the problem size grows, demonstrating its clear advantage in large-scale vehicular networks. 
\end{abstract}


\IEEEpeerreviewmaketitle

%\footnote{Please include the affiliations of all the authors as in a proper journal submission. Thanks.}

\section{Introduction} \label{sec:introduction}
Due to limited computation and storage capabilities of vehicular users, it is rather difficult to meet the strict requirements of Intelligent Transportation System (ITS) applications, i.e., intensive computation and content caching with low latency \cite{LTTan2018,Silva2021}. To this end, Vehicular Edge Computing (VEC), as an application of Mobile Edge Computing (MEC) in high-mobility environments, has been proposed as a solution \cite{ZNing2020,Girma2020,ChangHuan2022,YiMeng2022}. Even so, there remains the significant challenge to efficiently allocate the limited communication and computation resources of servers in VEC, due to an increasing number of vehicles that need their tasks processed.

\begin{figure*}[!tbp]
        \centering
        \includegraphics[width=2\columnwidth]{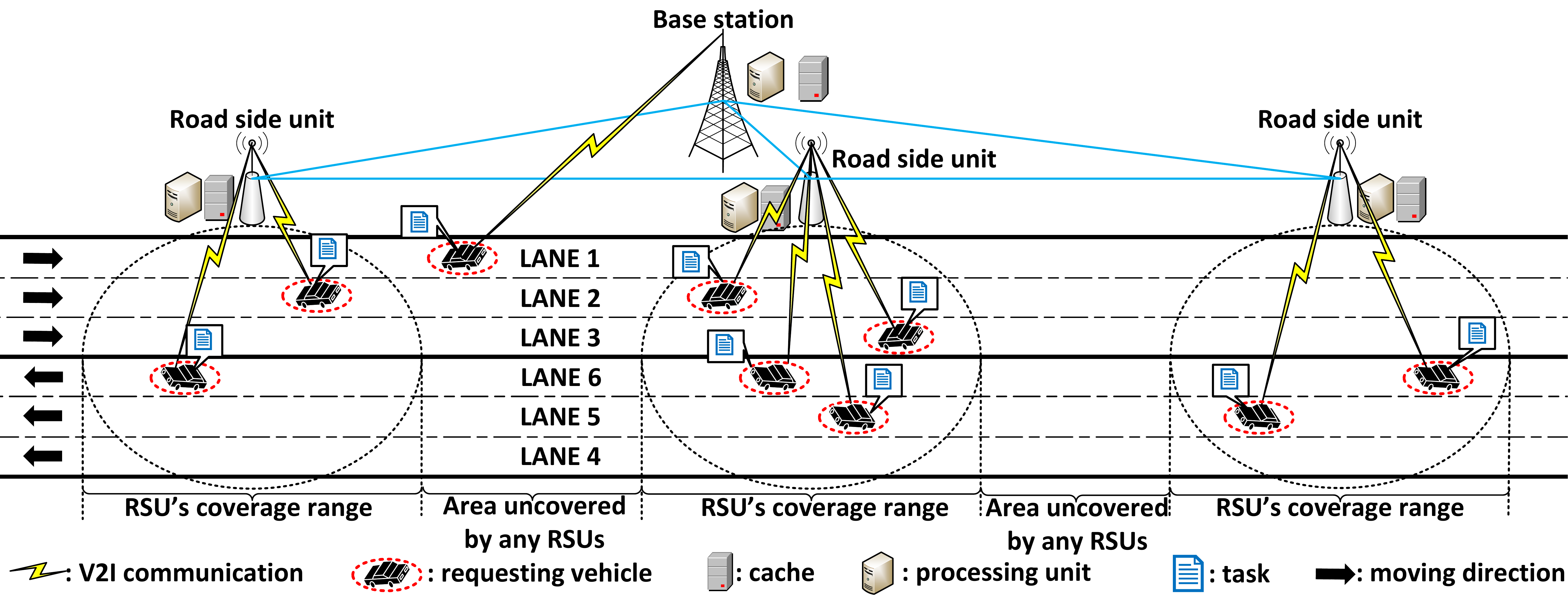}
        \captionsetup{justification=centering}
        \caption{Task assignment in VEC-based ITS applications.}
        \label{fig:scenario}
        \vspace{2mm}
\end{figure*}

We consider network scenarios as depicted in Fig.~\ref{fig:scenario}. Specifically, autonomous vehicles left-drive in two directions along a six-lane highway, similar to M$1$ Pacific Highway linking Newcastle with Sydney in New South Wales, Australia \cite{NSW2019}. A macrocell base station (BS) is deployed to provide network connectivity along the highway. For data and computation offloading, a set $\mathcal{R}$ of road side units (RSUs) are deployed at an inter-RSU distance of $D_{\text{R}}$ to bring the network closer to the vehicles. We denote by $\{\text{BS}\}$ the set containing only the BS, and by $|\mathcal{R}|$ the number of RSUs. The BS and RSUs are connected via wired links for load balancing and control coordination. Each of them is equipped with a server comprising a data processing unit and a cache.

Let us assume that the vehicles have to complete computation-intensive tasks. Due to their limited computing resources, it is sensible to offload these tasks to the servers at the BS and/or the RSUs. The offload requests are sent via vehicle-to-infrastructure (V2I) communication, which is supported by the Long-Term Evolution-Advanced (LTE-A) protocol. We denote by $\mathcal{I}$ the set of requesting vehicles, and assume a vehicle only requests to offload one task at a time. As such, we refer to vehicle $i \in \mathcal{I}$ and task $i \in \mathcal{I}$ interchangeably. Also, the number of tasks to be offloaded is equal to the number of requesting vehicles. 

If a vehicle $i \in \mathcal{I}$ traveling at a constant speed of $v_i$ is still within the coverage range of an RSU $r \in \mathcal{R}$, its offload request is sent directly to the RSU $r$; otherwise, to the BS. In either case, the server at the BS collects from all the RSUs information about task sizes, server computing capabilities, and current location and speed of vehicles. It then computes and makes a task assignment decision as to where the tasks are to be processed, i.e., at the BS or the RSUs; and in the latter case, which RSU in $\mathcal{R}$.

\subsection{Background}

To address the issue of inadequate provisioning of computation resources for multiple users, \cite{Bi2021} proposes an algorithm that optimally distributes tasks from smart mobile devices (SMDs) to MEC servers. By using a combination of particle swarm optimization, simulated annealing and genetic algorithms, this approach minimizes the energy consumed by SMDs and servers while also optimizing the task offloading ratio. In a related work by \cite{Yuan2021}, tasks sent from the computers and iPads in the terminal layer are allocated to servers in the edge computing layer and cloud data layer. In order to maximize the total network profit (which is the net revenue discounted by a cost), the authors propose a task allocation strategy that utilises a %migrating-birds optimization 
swarm intelligence 
approach based on simulated annealing. However, without taking the mobility of SMD users into account in the problem formulation, these two algorithms are only applicable to static environments, rather than high-mobility environments (i.e. in ITS or vehicular networks).

%Without taking SMD users' mobility in the problem formulation, this algorithm could be only applicable in static environments. 

%\cite{Yuan2021} allocates task sent from ipad and computers in terminal layers to servers in edge computing layer and cloud data center layer. With the target of maximizing the total network profit, the proposed algorithm is relied on simulated-annealing-based migrating birds optimization. 

%This algorithm is able to assign the tasks to the servers in edge computing and CDC layers optimally in static environments instead high mobility environments. 

Unlike \cite{Bi2021,Yuan2021}, the work of \cite{Zeng2021} develops a task assignment algorithm where tasks requested by vehicles are assigned to either VEC servers or volunteer vehicles in a vehicular network. The developed algorithm is based on a Stackelberg game where VEC servers and requesting vehicles are respectively modelled as leaders and followers. To completely process all the tasks, the servers recruit more volunteers while setting and sending prices to the requesting vehicles. The game strategy is to 1) maximize the income of VEC servers and volunteer vehicles, 2) reduce the cost incurred to the servers and volunteers, and 3) minimize the payment made by the requesting vehicles for processing their tasks. However, when the requesting vehicles and volunteer vehicles move in different directions and at different speeds, their connection time is limited to a brief amount due to the short communication range of vehicles (about 300 m). As a result, the requesting vehicles will be out-of-range of the volunteer vehicles, while the latter have not completely processed the tasks requested by the former.

%\cite{Bai2021}: the objective is to visit as many targets as possible using the minimum number of vehicles with a minimum travelling time. The application of this paper is quite different from that of our paper. 

%\cite{Zhang2020}: proposes three construction-based heuristic methods to assign pairings to individual crew members to form rosters. This application is not similar to our considered application. 

The optimization methods for resource allocation in \cite{Bi2021,Yuan2021,Bai2021,Zhang2020,Zeng2021} require accurate knowledge of channel conditions which are typically time-varying and, oftentimes, unavailable in high-mobility scenarios. These solutions are typically based on a snapshot model of the vehicular networks, while ignoring the long-term influence of the current decision \cite{ZNing2020}. By contrast, without any prior knowledge of the operating environment, reinforcement learning (RL) is able to make decisions that maximize the long-term rewards for the networks according to \cite{zong2022mapdp} and \cite{sarkar2022}. It is arguably a promising tool to tackle problems encountered in task offloading, and communication and computation resource allocation in VEC-based ITS with time varying and unknown channel conditions.

In \cite{WZhan2020}, multiple in-car applications employ an RL-based scheduling strategy to offload their tasks to MEC servers located within road side units (RSUs). Here, the latency and energy consumption for task processing are minimized. Taking a step further, a joint management scheme of spectrum, computing and storing resources in VEC is proposed in \cite{HPeng2020} using deep reinforcement learning (DRL). Note that in \cite{WZhan2020} and \cite{HPeng2020}, vehicles potentially reside within the coverage range of RSUs for a short time duration only, due to their high mobility and the limited communication range of the serving RSUs (about 600 m); hence, it is possible that a vehicle moves out of the range of its serving RSU even before that RSU processes its tasks completely.
%its tasks are completely processed.

The above issue can be overcome by allowing the vehicle to migrate its tasks to the MEC servers of the next RSUs that the vehicle is about to move into. For example, in \cite{HWang2020}, there is an autonomous vehicle moving along a highway or a city expressway, and its tasks are migrated between MEC servers and processed. Assuming these MEC servers have large computation resources, \cite{HWang2020} use DRL to minimize the energy consumption for task processing while meeting latency requirements. In addition, only a \emph{single} agent interacts with the environment to determine an optimal task migration policy. Also based on DRL, \cite{QYuan2020} not only develop a task migration scheme but also find the best migration routes for vehicles in \emph{urban} areas. Here, a vehicle only migrates its tasks to an MEC server if the time it takes that vehicle to reach such a server is the shortest. Compared to \cite{HWang2020}, the work of \cite{QYuan2020} could be applicable to multi-agent systems owing to utilizing communication and cooperation between multiple autonomous vehicles. However, this scheme might cause a change in the original route of vehicles as the tasks are not migrated with respect to the vehicles' mobility pattern.

% The task migration algorithm \cite{HWang2020} or the joint task migration and mobility optimization algorithm \cite{QYuan2020} is based on DRL.

A major issue with the DRL approaches in \cite{HWang2020} and \cite{QYuan2020} is that a significant training time is required in large environments, e.g. $100$ or more vehicles/agents, and the algorithm's convergence is not guaranteed. To address this issue, \cite{DDNguyen2017} and \cite{CFan2020} employ regret matching (RM) learning to design algorithms for multi-agent systems. The advantages of RM learning-based algorithms in several applications, e.g. seasonal forecasting and learning in matching markets without incentives, have been demonstrated by \cite{Xu2020,Flore2021,Genevieve2021,Bistritz2020}. In particular, these algorithms can converge to correlated-equilibrium solutions faster than RL-based algorithms as shown in \cite{Daskalakis2022} and \cite{Ioannis2022}. Additionally, it is unnecessary that the correlated equilibrium solutions must be the optimal solution. Their algorithms, however, are not specifically designed for task migration in VEC, and their solutions may be rendered inapplicable due to the inherited characteristics of vehicle networks with high mobility.

\subsection{Contributions}

In this paper, we propose a RM learning-based task assignment scheme that minimizes the total delay and cost incurred by vehicles in a \emph{highway} scenario like \cite{HWang2020}. We assume that once a vehicle leaves the coverage area of its serving RSU, it will migrate its tasks to other suitable RSUs or a macrocell base station according to its mobility pattern. The contributions of this paper are summarized as follows.
\begin{enumerate}
  \item To improve over \cite{WZhan2020} and \cite{HPeng2020}, we formulate a task assignment problem as a binary nonlinear programming (BNLP) problem with specific constraints on the movement of participating vehicles. Compared to \cite{HWang2020} and \cite{QYuan2020}, this problem is formulated for migrating the tasks of \emph{multiple} autonomous vehicles between servers according to these vehicles' movement.
  
  \item Unlike \cite{HWang2020} and \cite{QYuan2020}, we reformulate the BNLP problem as a standard repeated game. Then, we propose a distributed RM algorithm that decomposes the state observations and actions of a monolithic centralized agent into those of multiple agents. In particular, this iterative game-based learning algorithm is able to guarantee an equilibrium solution. We further propose a forgetting method to speed up the convergence of the traditional RM algorithms in \cite{DDNguyen2017,CFan2020}. Doing so allows the algorithm to effectively handle the high level of user mobility in vehicle networks. % We prove theoretically that the proposed algorithm guarantees convergence to the set of CE solutions that contain the optimal solutions to the BNLP problem. 
  
  \item Our simulation results with practical parameter settings demonstrate the advantages of our solution in terms of delay and cost minimization, utility fairness among agents, and convergence speed particularly in large-scale network settings.% The results also confirm that shortening the inter-RSU distance further improves the network performance.
\end{enumerate}

The remainder of the paper is organized as follows. Section~\ref{sec:system model} presents the system model, including a wireless communication model and a computation model, while Section~\ref{sec:problem_formulation} describes the problem formulation for task assignment. Then, Section~\ref{sec:proposed scheme} proposes an RM-based solution to the task assignment problem. Here, the problem is reformulated as a repeated game while the definition of a correlated equilibrium is introduced. Section~\ref{sec:simulation results} conducts simulations to illustrate the efficiency of the proposed approach. Finally, we summarize the paper in Section~\ref{sec:conclusion}.

\section{System Model} \label{sec:system model}

In our scenarios, once a requesting vehicle wants its task to be processed by a server at an RSU or a BS, it must send the task to the RSU/BS via a wireless link. In addition, the task can be migrated from the RSU/BS to the others via a wired connection. Thus, we first model wireless communication between the requesting vehicles and the RSUs/BS in Section~\ref{sec:communication model}. Then, to determine the amount of time and cost needed for 1) uploading tasks through wireless links, 2) migrating tasks between RSUs/BS through wired links, and 3) processing tasks completely, we develop a computation model in Section~\ref{sec:computation model}. The delay and cost will be used for our problem formulation in Section~\ref{sec:problem_formulation}.

\subsection{Communication Model} \label{sec:communication model}

We consider that the received signal strength at the RSUs and BS depends only on the positional shift of the vehicles, where the effect of small-scale fading is averaged out. For interference cancellation, we adopt the orthogonal frequency-division multiplexing (OFDM) to assign orthogonal frequencies to the link between an RSU/BS $r \in \mathcal{R} \cup \{\text{BS}\}$ and a vehicle $i\in\mathcal{I}$. The data rate at which the tasks of the vehicle $i$ are uploaded to the RSU/BS $r$ at a given time $t$ is expressed as:
\begin{equation}
    R_{r,i}(t) = B_{r,i}(t)\log_2\bigg(1+\frac{p_i|h_{r,i}(t)|^2}{{N_r}^2}\bigg) \ \
    \substack{\forall i \in \mathcal{I},\\ \forall r \in \mathcal{R} \cup \{\text{BS}\}},
\end{equation}
where $B_{r,i}(t)$ is the link's bandwidth, $p_i$ is the transmit power of the vehicle $i$, $|h_{r,i}(t)|^2$ is the link gain between the vehicle $i$ and the RSU/BS $r$, and ${N_r}^2$ is the received noise power. %Because the link gain depends on the distance between vehicle $i$ and RSU/BS $r$, it holds that\footnote{I think you have deleted this equation!!!}. 
Here, $|h_{r,i}(t)|^2=f(d_{r,i}(t))$ with $f(.)$ a path-loss function, and $d_{r,i}(t)$ the Euclidean distance between the vehicle $i$ and the RSU/BS $r$ at the time $t$.

\subsection{Computation Model} \label{sec:computation model}
%\subsection{Network model}
%Before a vehicle $i$'s task is completed at a server, it has to be uploaded to an RSU $r$ if vehicle $i$ resides within RSU $r$'s coverage range. Otherwise, this task is uploaded to the BS.
The amount of time needed for a task $i\in\mathcal{I}$ to be uploaded to an RSU/BS $r \in \mathcal{R} \cup \{\text{BS}\}$ is given by:
\begin{equation}
    T_{r,i}^{\text{u}}(t)=\frac{s_i}{R_{r,i}(t)} \ \
    \substack{\forall i \in \mathcal{I},\\ \forall r \in \mathcal{R} \cup \{\text{BS}\}},
    \label{eq:upload time}
\end{equation}
where $s_i$ is the size of the task $i$.

We use two binary variables $x_{r,i}(t)$ and $x_{r,\hat{r},i}(t)$ to decide where the task $i\in\mathcal{I}$ is executed at the time $t$. If the task $i$ is to be processed at an RSU/BS $r\in\mathcal{R}\cup \{\text{BS}\}$, then $x_{r,i}(t)=1$; otherwise, $x_{r,i}(t)=0$. If the task $i$ is migrated and processed at the other RSU/BS $\hat{r} \in \mathcal{R} \cup \{\text{BS}\} \setminus \{r\}$, then $x_{r,\hat{r},i}(t)=1$; otherwise, $x_{r,\hat{r},i}(t)=0$. The task migration time is calculated as \cite{QYuan2020}:
\begin{equation}
    T_{r,\hat{r},i}^{\text{m}}(t)= x_{r,\hat{r},i}(t)\bigg(\frac{s_i}{B_{\text{R}}} + 2 \cdot \alpha \cdot h_{r,\hat{r}} \bigg) \ \
    \substack{\forall i \in \mathcal{I},\\ \forall r \in \mathcal{R} \cup \{\text{BS}\},\\ \forall \hat{r} \in \mathcal{R} \cup \{\text{BS}\} \setminus \{r\} },
    \label{eq:mirgation time}
\end{equation}
where $B_{\text{R}}$ is the bandwidth of the wired link between $r$ and $\hat{r}$, $\alpha$ is the coefficient of migration delay, and $h_{r,\hat{r}}$ is the number of hops between $r$ and $\hat{r}$.

The processing delay for the task $i$ is calculated as:
\begin{equation}
T_{r,\hat{r},i}^{\text{p}}(t) = \frac{x_{r,i}(t)\cdot f_i}{F_{r,i}} + \frac{x_{r,\hat{r},i}(t) \cdot f_i}{F_{\hat{r},i}} \ \
    \substack{\forall i \in \mathcal{I},\\ \forall r \in \mathcal{R} \cup \{\text{BS}\},\\ \forall \hat{r} \in \mathcal{R} \cup \{\text{BS}\} \setminus \{r\}},
    \label{eq:processing time}
\end{equation}
where $f_i$ is the number of CPU cycles required to completely process the task $i$, and $F_{r,i}$ and $F_{\hat{r},i}$ cycles/s are respectively the computation capacity allocated to the task $i$ by $r$ and $\hat{r}$.

From Eqs.~\eqref{eq:upload time},~\eqref{eq:mirgation time} and~\eqref{eq:processing time}, the total delay to complete the task $i$ is calculated as:
\begin{equation}
    \begin{aligned}
    T_{r,\hat{r},i}^{\text{}}(t)=& T_{r,i}^{\text{u}}(t)+T_{r,\hat{r},i}^{\text{m}}(t)+T_{r,\hat{r},i}^{\text{p}}(t).%\\
    %=&\frac{s_i}{R_{r,i}(t)}+x_{r,\hat{r},i}(t)\bigg(\frac{s_i}{B_{\text{R}}} + 2 \cdot \alpha \cdot h_{r,\hat{r}} \bigg)
    %\\&+\frac{x_{r,i}(t)\cdot f_i}{F_{r,i}}+ \frac{x_{r,\hat{r},i}(t) \cdot f_i}{F_{\hat{r},i}} \ \
    %\substack{\forall i \in \mathcal{I},\\ \forall r \in \mathcal{R} \cup \{\text{BS}\},\\ \forall \hat{r} \in \mathcal{R} \cup \{\text{BS}\} \setminus \{r\}}.
    \end{aligned}
    \label{eq:total delay}
\end{equation}

%\subsection{Cost Analysis of Offloading Tasks}

%\noindent

\noindent Similar to \eqref{eq:total delay}, the cost for processing task $i$ is given by:
\begin{equation}
    c_{r,\hat{r},i}(t) = c_{r,i}^{\text{u}}(t) + c_{r,\hat{r},i}^{\text{m}}(t) + c_{r,\hat{r},i}^{\text{p}}(t)\ \
    \substack{\forall i \in \mathcal{I},\\ \forall r \in \mathcal{R} \cup \{\text{BS}\},\\ \forall \hat{r} \in \mathcal{R} \cup \{\text{BS}\} \setminus \{r\}},
    \label{eq:total cost}
\end{equation}
where $c_{r,i}^{\text{u}}$(t), $c_{r,\hat{r},i}^{\text{m}}(t)$ and $c_{r,\hat{r},i}^{\text{p}}(t)$ are respectively the costs of task uploading, task migrating and task processing. 

Specifically,
\begin{equation}
  c_{r,i}^{\text{u}}(t)=\delta^{r}_{\text{u}} \cdot B_{r,i}(t) \ \
    \substack{\forall i \in \mathcal{I},\\ \forall r \in \mathcal{R} \cup \{\text{BS}\}},
  \label{eq:upload cost}
\end{equation}
where $\delta^r_{\text{u}}>0$ is the communication cost.

After the task $i$ is uploaded to $r$, a service entity hosted at $r$ will handle the task $i$. This entity is migrated from $r$ to $\hat{r}\in \mathcal{R} \cup \{\text{BS}\} \setminus \{r\}$ if the task $i$ is not completely processed before the vehicle $i$ leaves the coverage area of $r$. To migrate the service entity from $r$ to $\hat{r}$, the vehicle $i$ incurs the following cost \cite{QYuan2020}:
\begin{equation}
  c_{r,\hat{r},i}^{\text{m}}(t)= x_{r,\hat{r},i}(t)\cdot \delta^{r,\hat{r}}_{\text{m}} \cdot \theta \ \
    \substack{\forall i \in \mathcal{I},\\ \forall r \in \mathcal{R} \cup \{\text{BS}\},\\ \forall \hat{r} \in \mathcal{R} \cup \{\text{BS}\} \setminus \{r\}},
  \label{eq:migration cost}
\end{equation}
where $\delta^{r,\hat{r}}_m>0$ is the migration cost and $\theta$ is the data size of each service entity.

The computation cost for the task $i$ at either RSU/BS $r$ or $\hat{r}$ is expressed as:
\begin{equation}
  c_{r,\hat{r},i}^{\text{p}}= x_{r,i}(t)\cdot\delta^r_{\text{p}}\cdot F_{r,i} + x_{r,\hat{r},i}(t) \cdot \delta^{\hat{r}}_{\text{p}} \cdot F_{\hat{r},i} \ \
    \substack{\forall i \in \mathcal{I},\\ \forall r \in \mathcal{R} \cup \{\text{BS}\},\\ \forall \hat{r} \in \mathcal{R} \cup \{\text{BS}\} \setminus \{r\}},
  \label{eq:computation cost}
\end{equation}
where $\delta^r_{\text{p}}>0$ and $ \delta^{\hat{r}}_{\text{p}}>0$ are the unit computation costs. 

%From~\eqref{eq:cost},~\eqref{eq:upload cost},~\eqref{eq:migration cost} and~\eqref{eq:computation cost},
%\begin{equation}
%  \begin{aligned}
%  c_{r,\hat{r},i}(t)=&\delta^{r}_{\text{u}} \cdot B_{r,i}(t)
%  +x_{r,\hat{r},i}(t) \cdot \delta^{r,\hat{r}}_{\text{m}} \cdot \theta
%  +x_{r,i}(t)\cdot\delta^r_{\text{p}}\cdot F_{r,i} \\
%  &+ x_{r,\hat{r},i}(t) \cdot \delta^{\hat{r}}_{\text{p}} \cdot F_{\hat{r},i} \ \
%    \substack{\forall i \in \mathcal{I},\\ \forall r \in \mathcal{R} \cup \{\text{BS}\},\\ \forall \hat{r} \in \mathcal{R} \cup \{\text{BS}\} \setminus \{r\}}.
%  \end{aligned}
%  \label{eq:total cost}
%\end{equation}

%{\color{blue} \textbf{Question:} What is the unit of each cost? Can the cost be 0 or negative?}

%{\color{blue} \textbf{Answer from Long:} I consider that the cost unit in this paper is $\$$ because the references cited in the paper have not mentioned the cost unit specifically. No, the cost has to be positive because it is the price to charge users. }

\section{Problem Formulation} \label{sec:problem_formulation}

There are three constraints that describe the dependence of task assignment on the vehicle mobility, different from \cite{QYuan2020}. When a task $i\in\mathcal{I}$ is completely executed by an RSU $r\in\mathcal{R}\cup\{\text{BS}\}$, the delay for completing the task $i$ must not be larger than the duration that the vehicle $i$ resides within $r$'s coverage area. As such,
\begin{equation}
x_{r,i}(t) \bigg[T_{r,\hat{r},i}^{\text{}}(t) - \frac{\tilde{d}_{r,i}(t)}{v_i}\bigg] \leq 0 \ \
\substack{\forall i \in \mathcal{I},\\ \forall r \in \mathcal{R},\\\forall \hat{r} \in \mathcal{R} \setminus \{r\}},
\label{eq:vehicular 1}
\end{equation}
where $\tilde{d}_{r,i}(t)$ is the distance that the vehicle $i$ travels before leaving the coverage area of $r$.

If the task $i$ is migrated from the RSU $r\in\mathcal{R}$ to another RSU $\hat{r}\in \mathcal{R} \setminus \{r\}$, the delay is instead constrained by:
\begin{equation}
x_{r,\hat{r},i}(t)\bigg[T_{r,\hat{r},i}^{\text{}}(t)-\frac{\tilde{d}_{r,i}(t)+D_{\text{R}} \cdot h_{r,\hat{r}}}{v_i}\bigg] \leq 0 \ \ \substack{\forall i \in \mathcal{I},\\ \forall r \in \mathcal{R},\\\forall \hat{r} \in \mathcal{R} \setminus \{r\}}.
\label{eq:vehicular 2}
\end{equation}

If the task $i$ is migrated from the BS to an RSU $\hat{r}\in\mathcal{R}$, the delay is then constrained by:
\begin{equation}
  \begin{aligned}
  & x_{r,\hat{r},i}(t)\bigg[T_{r,\hat{r},i}^{\text{}}(t)-\frac{\tilde{d}_{r,\breve{r},i}(t)+2R_{\text{R}}+D_{\text{R}}\cdot h_{\breve{r},\hat{r}}}{v_i}\bigg] \leq 0\\
  &\forall i \in \mathcal{I}, r\in \{\text{BS}\},\forall \breve{r},\hat{r} \in \mathcal{R},
  \end{aligned}
  \label{eq:vehicular 3}
\end{equation}
where  $\tilde{d}_{r,\breve{r},i}$ is the distance that the vehicle $i$ has traveled in the area uncovered by any RSUs before it enters the coverage area of the closest RSU $\breve{r}$, and $R_{\text{R}}$ is the communication range of an RSU.

We aim to minimize the total delay and cost for completing all $|\mathcal{I}|$ tasks. The task assignment in vehicular edge computing is thus formulated as the following BNLP problem.
\begin{subequations}
    \label{main problem}
    \begin{alignat}{2}
        \min_{\substack{x_{r,i}(t)\\ x_{r,\hat{r},i}(t)}}\quad
        \begin{split}
        &\sum_{i \in \mathcal{I}}\displaystyle\sum_{\substack{r,\hat{r} \in \mathcal{R}\cup\{\text{BS}\}\\ r \neq \hat{r}}} \Big[\beta \cdot T_{r,\hat{r},i}^{\text{}}(t) +\gamma \cdot c_{r,\hat{r},i}(t)\Big]
        \end{split} \label{objective function}\\
        \text{s.t.}\quad &\eqref{eq:total delay},\eqref{eq:total cost},\eqref{eq:vehicular 1},\eqref{eq:vehicular 2},\eqref{eq:vehicular 3}\\
                         & \sum_{i \in \mathcal{I}} x_{r,i}(t) \cdot F_{r,i} \leq F^{\max}_r \quad \forall r \in \mathcal{R}, \label{computation 1}\\
                         \begin{split}
                         & \sum_{i \in \mathcal{I}}\sum_{r\in\mathcal{R}} x_{r,\hat{r},i}(t) \cdot F_{\hat{r},i} \leq F^{\max}_{\hat{r}}\\ & \hspace{0.5cm} \forall \hat{r} \in \mathcal{R} \cup \{\text{BS}\}\setminus \{r\},
                         \end{split} \label{computation 2}\\
                         \begin{split}
                         & \sum_{i \in \mathcal{I}} x_{r,i}(t)\cdot F_{r,i}+\sum_{\hat{i} \in \mathcal{I} \setminus \{i\}}\sum_{\hat{r} \in \mathcal{R}\setminus \{r\}} x_{\hat{r},r,\hat{i}}(t)\\
                         & \times F_{r,\hat{i}} \leq F^{\max}_r \quad \forall r \in \mathcal{R},
                         \end{split} \label{computation 3}\\
                         \begin{split}
                         & x_{r,i}(t) + \sum_{\hat{r} \in \mathcal{R} \cup \{\text{BS}\}\setminus \{r\}} x_{r,\hat{r},i}(t) = 1\\
                         & \hspace{0.5cm} \forall r \in \mathcal{R},\forall i \in \mathcal{I},
                         \end{split} \label{task allocation}\\
                         \begin{split}
                         & x_{r,i}(t),x_{r,\hat{r},i}(t) \in \{0,1\} \quad \forall i \in \mathcal{I},\\
                         &\hspace{0.5cm} \forall r \in \mathcal{R}\cup \{\text{BS}\}, \forall \hat{r} \in \mathcal{R} \cup \{\text{BS}\}\setminus \{r\},
                         \end{split}
    \end{alignat}
\end{subequations}
where $\beta>0$ and $\gamma>0$ are the weights to prioritize the delay and the cost, respectively. %dditionally, the maximum computing capabilities of the server at RSUs/BS $r$ and $\hat{r}$ are denoted as $F^{\max}_r$ and $F^{\max}_{\hat{r}}$, respectively.
Constraints~\eqref{computation 1},~\eqref{computation 2} and~\eqref{computation 3} show that the computation capacity assigned to a task $i$ is upper-bounded by the maximum computation capacities $F^{\max}_r$ or $F^{\max}_{\hat{r}}$. Constraint~\eqref{task allocation} enforces that an arbitrary task $i$ must be processed by one of the RSUs and the BS.

%{\color{blue} \textbf{Question:} What are the units of $\beta$ and $\gamma$ in (14)? Do they have to be non-negative?}

%{\color{blue} \textbf{Answer from Long:} The unit of $\beta$ is $\frac{1}{second}$ while the unit of $\gamma$ is $\frac{1}{\$}$. No, they have to be positive because we want to minimize both delay and cost.}

\section{Proposed Multi-Agent Regret-Matching Learning based Task Assignment Scheme} \label{sec:proposed scheme}

\begin{algorithm*}[t]
    \caption{Multi-Agent RM Learning-Based Task Assignment Algorithm}
    \label{alg:RMalg}
    \begin{algorithmic}[1]
        \STATE \textbf{Initialization:} Each player $i$ initializes its action selection policy with a uniform strategy $\pi_i^{(1)}(j) \leftarrow \frac{1}{|\Ac_i|}$ $\forall j\in \Ac_i$
        \STATE \textbf{Main algorithm:} Each player $i\in\mathcal{I}$ independently runs the following procedure to decide its action over time
        \STATE \textbf{for} $t=1,2, \dots$ \textbf{do}
        \STATE \quad \textit{Action selection:} Player $i$ samples an action $a^{(t)}_i=j \in \mathcal{A}_i$ from its probability distribution of action selection $\pi_i^{(t)}$. The BS then updates the chosen action of player $i$ to all other players.
        \STATE \quad \textit{Utility update:} Player $i$ receives a utility as a result of its chosen action $u^{(t)}_i\left(j,a^{(t)}_{-i}\right)$ computed by Eq.~\eqref{eq:utility model}.
        \STATE \quad \textbf{for} $k\in \mathcal{A}_i\setminus\{j\}$
        \STATE \quad \quad \textit{Expected utilities:} Using Eq.~\eqref{eq:utility model}, player $i$ calculates an expected utility $u^{(t)}_i\left(k,a^{(t)}_{-i}\right)$ if choosing an action $k\neq j$, given the choices made by the other players.
        \STATE \quad \quad \textit{Regret update:} Using Eq.~\eqref{eq:regretmatching_recursive_forgetting}, player $i$ computes the cumulative regret $\bar{D}_i^{(t)}(j,k)$ for not choosing $k$.
        \STATE \quad \quad \textit{Strategy update:}  Using Eq.~\eqref{eq:CRM}, player $i$ updates its next action probability $\pi_i^{(t+1)}(k)$.
        \STATE \quad \textbf{end for}
        \STATE \quad Player $i$ plays the same action chosen in the previous round with the remaining probability\\ \qquad \qquad $\pi_i^{(t+1)}(j)=1-\displaystyle\sum\nolimits_{k\neq j}\pi_i^{(t+1)}(k)$.
        \STATE \textbf{end for}
    \end{algorithmic}
    %\vspace{-0.1cm}
\end{algorithm*}

The optimization problem in \eqref{main problem} is nonconvex and combinatorial with nonlinear constraints. Traditional optimization methods may not be able to return a solution within an acceptable time frame, which is an important requirement in vehicular networks with a high degree of mobility. As such, we propose an iterative game-based learning algorithm that guarantees an equilibrium solution. The proposed algorithm is based on the regret minimization procedure~\cite{hart2000simple, Hart_2001}. This procedure is well-known for its low complexity and provable convergence when making decisions in a situation involving multiple stakeholders.

In this paper, we consider that all tasks are homogeneous (see Section~\ref{sec:introduction}), and the number of tasks is predefined (see Section~\ref{sec:problem_formulation}). The proposed solution is readily applicable to scenarios where tasks have different deadlines or different sizes, or the number of tasks varies over time. We further introduce a forgetting factor in the learning algorithm to enable fast convergence---an essential requirement in a fast-changing environment due to highly-mobile learning agents (i.e., vehicles). Using simulations with realistic network settings, we will later show that our solution adapts and converges much faster than existing approaches, especially as the number of participating vehicles (i.e., tasks) increases. 

\subsection{Game Reformulation for Task Assignment}
We propose to reformulate problem~\eqref{main problem} as a multi-agent distributed learning problem. Here, each requesting vehicle is an independent decision maker who learns to jointly reach an optimal solution. To ensure convergence of the learning solution at the optimum point for all the requesting vehicles, we designate the BS as a central operator. After all the requesting vehicles have played their respective actions, the operator updates each vehicle with the actions chosen by the others.

Specifically, we model the task assignment in vehicular edge computing~\eqref{main problem} as a repeated game $\mathcal{G}=(\mathcal{I},\mathcal{A},\mathcal{U})$, where the players aim to minimize the long-run average delay and the cost to process the tasks presented by the requesting vehicles. In this model, the (finite) set of requesting vehicles $\mathcal{I}=\{1,2,\dots,|\mathcal{I}|\}$ is regarded as the set of players. Each player $i\in\mathcal{I}$ has its set of finite actions $\Ac_i = \Rc \cup \{\text{BS}\}$ as it decides where to offload its task to. We denote by $\mathcal{A}=\Ac_1\times \Ac_2\times \dots \times \Ac_{|\mathcal{I}|}$ the set of joint actions of all players. Let $\mathcal{U}=\{u_1,u_2,\dots,u_{|\mathcal{I}|}\}$ denote the set of utility functions of all the players.

To minimize the delay and cost for processing the tasks for the vehicles, the utility function of a player $i \in \mathcal{I}$ at a time $t$ resulting from a given action $a_i^{(t)}=r \in \Ac_i$ is designed as:
\begin{equation}
    u^{(t)}_i\left(a_i^{(t)},a^{(t)}_{-i}\right) = - \left[ \beta \cdot T^{\text{}}_{a_i^{(t)}|a_{-i}^{(t)}}(t) + \gamma \cdot c_{a_i^{(t)}|a_{-i}^{(t)}}(t)\right], 
    \label{eq:utility model}
\end{equation}
where $a^{(t)}_{-i}$ denotes the vector of RSU/BS actions decided by all the other $|\mathcal{I}|-1$ players at the time $t$. Here, if action $a_i^{(t)}$ of player $i$ satisfies all constraints in problem~\eqref{main problem}, we calculate the parameters $T^{\text{}}_{a_i^{(t)}|a_{-i}^{(t)}}(t)$ and $c_{a_i^{(t)}|a_{-i}^{(t)}}(t)$ using \eqref{eq:total delay} and~\eqref{eq:total cost}, respectively. Otherwise, $T^{\text{}}_{a_i^{(t)}|a_{-i}^{(t)}}(t) = \infty$ and $c_{a_i^{(t)}|a_{-i}^{(t)}}(t) = \infty$ where $\infty$ stands for a large positive value pre-assigned. Under this utility model, each player $i$ obtains a player-specific payoff depending on the joint action profile $\left(a_i^{(t)},a^{(t)}_{-i}\right)$ over all players. Here, maximizing the sum of all the players' utilities would result in minimizing the objective function in problem~\eqref{main problem}.

\subsection{Definition of Correlated Equilibrium}

In most cases, a game-based solution guarantees convergence to a set of equilibria, in which any vehicle does not achieve any gain by unilaterally changing their decision. It can be shown that the equilibrium of the reformulated game $\mathcal{G}$ is a correlated equilibrium (CE) \cite{Aumann_1987, Hart_1989}. %CE is an optimality concept introduced by Aumann~\cite{Aumann_1987} and is proven to exist for any finite games with bounded payoffs~\cite{Hart_1989}. 
%A CE models possible correlation between the players' actions compared to the usual strategic equilibrium of Nash where all players act independently.
A probability distribution $\psi$ defined on $\mathcal{A}$ is said to be a CE if for all player $i\in \mathcal{I}$, for all $a_{-i} \in \mathcal{A}_{-i}$ and for every pair of action $j,k \in \mathcal{A}_{i}$, it holds true that%\footnote{We write $\sum\nolimits_{s\in \mathcal{S}:i=j}$ for the sum over all $s$ in $\mathcal{S}$ whose $i$ equals $j$. Similar notations are used elsewhere in the paper.}
\begin{equation} 
\label{eq:correlatedequilibrium}
\sum_{a_{-i} \in \mathcal{A}_{-i}} \ \psi(j,a_{-i})\Big[u_i(k,a_{-i})-u_i(j,a_{-i})\Big]\leq0.
\end{equation}
When in a CE, each player does not benefit from choosing any other probability distribution over its actions, provided that all the other players do likewise. %When each player chooses their action independently of the other players, or without any implicit co-ordination mechanism, a CE is also a NE. 

\subsection{RM-based Learning with a Forgetting Factor}

An iterative algorithm that can be used to reach the CE set is the regret matching procedure proposed in~\cite{Hart_2001}. The key idea is to adjust the player's action probability to be proportional to the ``regrets'' for not having played other actions. Specifically, for any two actions $j \neq k \in \mathcal{A}_i$ at any time $t$, %given that $i_t=j$ is the action chosen by player A at time $t$, 
the cumulative regret of a player $i$ up to the time $t$ for not playing action $k$ instead of its actually played action $a_i^{(t)}=j$ is
%\begin{align}
%\bar{D}_i^{(t)} (j,k) = \displaystyle\frac{1}{t}\sum_{\tau=1}^t \mathbbm{I}{\big\{a_i^{(\tau)} = j\big\}} \left[u_i^{(\tau)}\left(k,a_{-i}^{(\tau)}\right)-u_i^{(\tau)}\left(j,a_{-i}^{(\tau)}\right)\right] \nonumber
%\end{align}
\begin{align}
\bar{D}_i^{(t)} (j,k) = \displaystyle\frac{1}{t}\sum_{\tau=1}^t \mathbbm{I}{\big\{a_i^{(\tau)} = j\big\}} \left[u_i^{(\tau)}\left(k,\cdot\right)-u_i^{(\tau)}\left(j,\cdot\right)\right], \nonumber
\end{align}
where $\mathbbm{I}(.)$ denotes the indicator function. This is the change in the average payoff that the player $i$  would have if choosing a different action $k\neq j$ every time they played $j$ in the past, given that all other players did not change their decisions. A positive value indicates a ``regret'' by the player $i$ for not having played action $k$ instead of the chosen action $j$. %at each iteration 

The regret can be recursively expressed as:
\begin{align}
\bar{D}_i^{(t)} (j,k) = \displaystyle \bigg(1-\frac{1}{t} \bigg) \bar{D}_i^{(t-1)} (j,k) + \frac{1}{t}\ D_i^{(t)} (j,k),
\label{eq:regretmatching_recursive}
\end{align}
where $D_i^{(t)} (j,k) =\mathbbm{I}{\{a_i^{(t)} = j\}} \left[u_i^{(t)}(k,a_{-i}^{(t)})-u_i(j,a_{-i}^{(t)})\right]$ denotes the instantaneous regret by the player $i$ for not playing the action $k$ instead of its played action $j$ at the time $t$.  Equation~\eqref{eq:regretmatching_recursive} updates the cumulative regret at each time step by adding a correction term based on the new instantaneous regret. As a stochastic approximation method, \eqref{eq:regretmatching_recursive}, although resulting in almost surely convergence, can be quite slow. This is especially true in a dynamic environment, where player's utility changes with time. This is likely to become a major issue in our considered vehicular networking scenario with a high degree of mobility.

\begin{figure}[t]
\centering
    \begin{minipage}[t]{0.5\textwidth}
    \centering
        \includegraphics[width=1.15\textwidth]{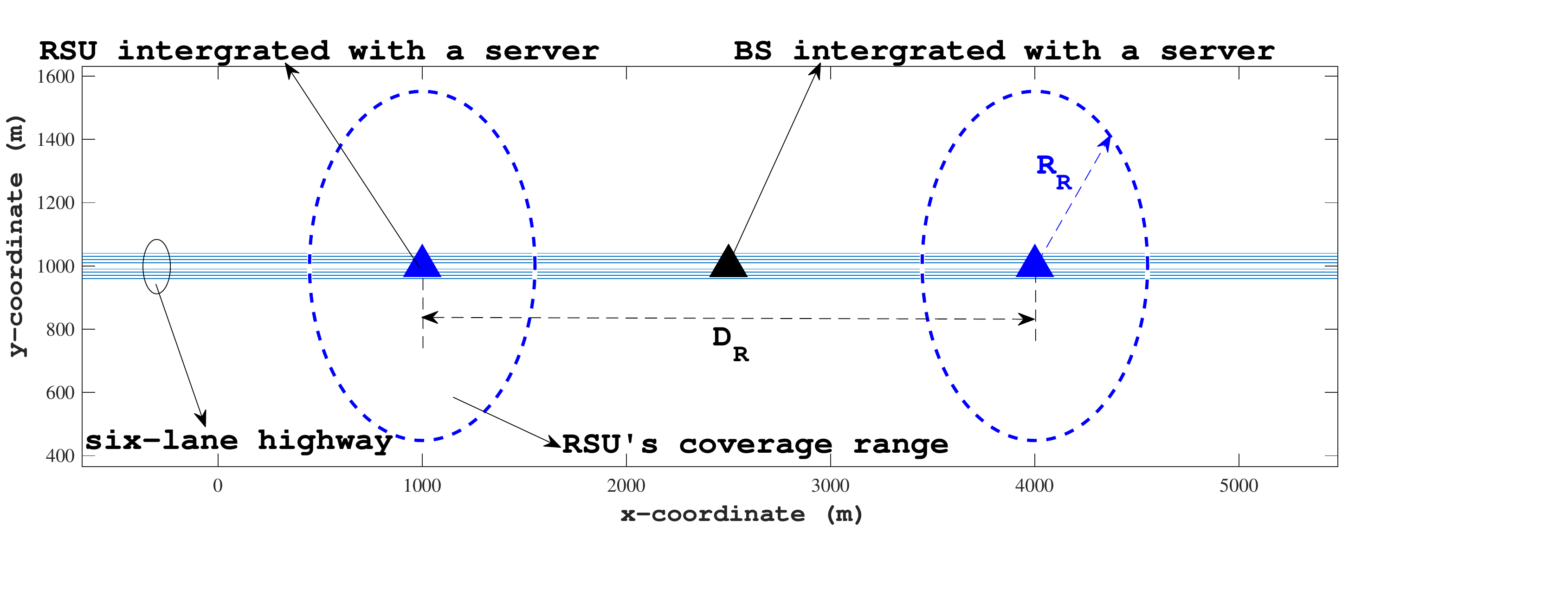}
        %\vspace{-1cm}
        \caption{{Deployment of $1$ BS and $2$ RSUs along a six-lane highway in Scenario~$1$.}}
        \label{fig:server_1}
    \end{minipage}
    
    \begin{minipage}[t]{0.5\textwidth}
    \centering
        \includegraphics[width=1.15\textwidth]{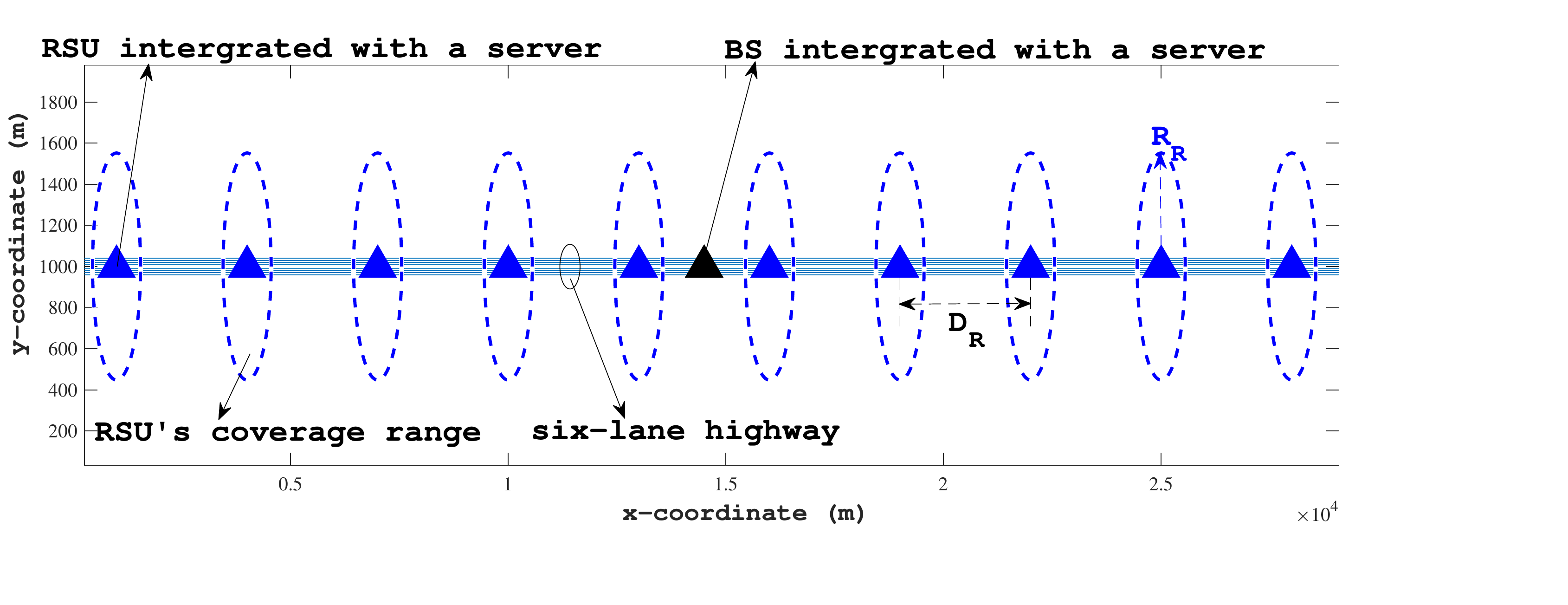}
        %\vspace{-1cm}
        \caption{{Deployment of $1$ BS and $10$ RSUs along a six-lane highway in Scenario~$2$.}}
        \label{fig:server_2}
    \end{minipage}
    
    \begin{minipage}[t]{0.5\textwidth}
    \centering
        \includegraphics[width=1.15\textwidth]{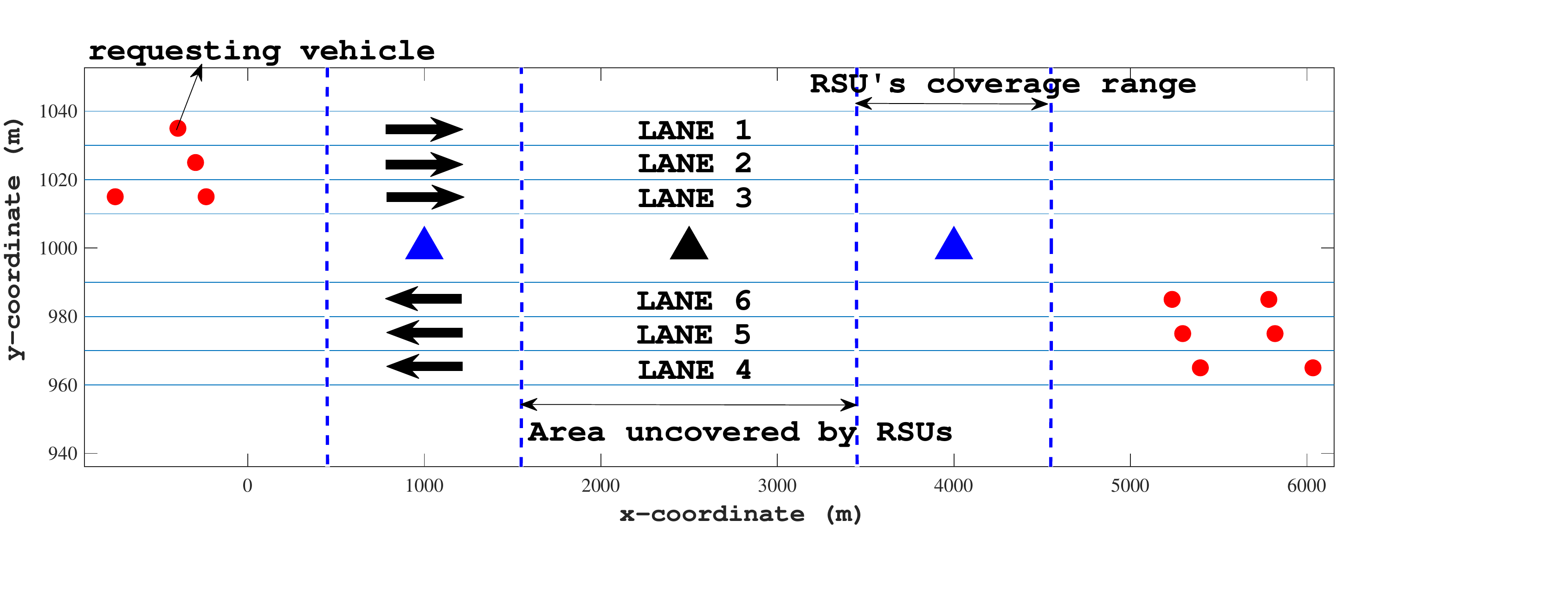}
        %\vspace{-1cm}
        \caption{{Snapshot of Scenario~$1$ ($3$ servers and $10$ requesting vehicles).}}
        \label{fig:scene_1}
    \end{minipage}

    \begin{minipage}[t]{0.5\textwidth}
    \centering
        \includegraphics[width=1.15\textwidth]{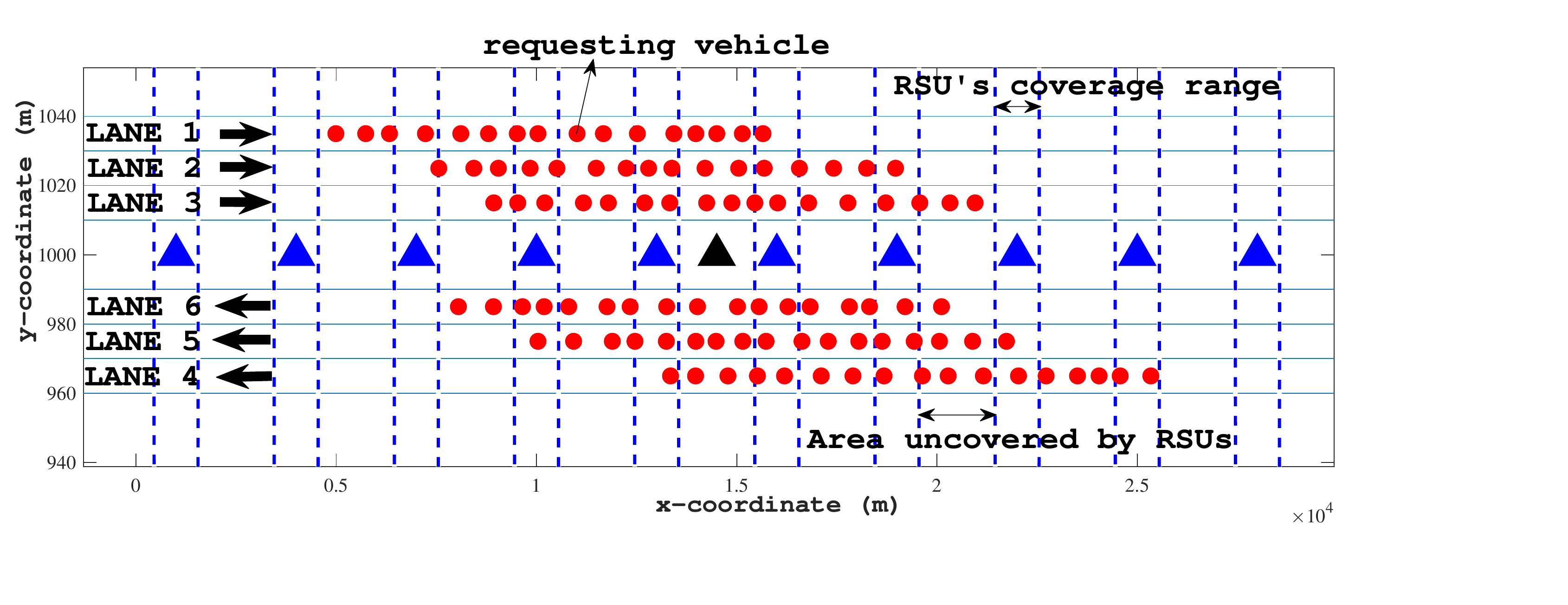}
       % \vspace{-1cm}
         \caption{{Snapshot of Scenario~$2$ ($11$ servers and $100$ requesting vehicles).}}
        \label{fig:scene_2}
    \end{minipage}
  % \vspace{-0.2cm}
\end{figure}

\begin{table*}[t]%\vspace{-0.1cm}
\caption{Performance comparison of the four algorithms in the two scenarios considered.}
%\vspace{-0.2cm}
\begin{center}
\setlength{\tabcolsep}{2pt}
%\begin{tabular}{|p{.8cm}|p{7.5cm}|m{2.1cm}|m{2.1cm}|m{2.1cm}|m{2.7cm}|}
\begin{tabular}{p{2.2cm}p{5.1cm}m{2.3cm}m{2.3cm}m{2.3cm}m{2.7cm}}
%\hline
%\cline{3-6}
\toprule
\multicolumn{1}{c}{}& \multicolumn{1}{c}{}&\textbf{ES} & \textbf{TRM} & \textbf{RLNF} & \textbf{Algorithm~\ref{alg:RMalg}} \\ [0.5ex]
\midrule %\hline \hline
\multirow{3}{*}{\hspace{2.5mm}\rotatebox{0}{\hspace{-0mm}\textbf{Scenario} $\mathbf{1}$}}%&{No. servers ($|\mathcal{R}|+1$)} & $3$ & $3$ & $3$ & $3$ \\
% & {No. servers RSUs ($|\mathcal{R}|$)}& $2$ & $2$ & $2$ & $2$ \\ 
% & {No. requesting vehicles (agents) ($|\mathcal{I}|$)} & $10$ & $10$ & $10$ & $10$ \\ 
% & {Vehicles' speed in lanes \{$1,4$\}, \{$2,5$\}, \{$3,6$\} (km/h) } & $90$, $100$, $120$ & $90$, $100$, $120$ & $90$, $100$, $120$ & $90$, $100$, $120$\\ 
% & {Inter-RSU distance ($D_{\text{R}}$) (km)} & 3 & 3 & 3 & 3 \\
 & Forgetting factor ($\lambda$)& N/A & N/A & N/A & $0.5$ \\
 & {Minimum sum of delay and cost} & ${\approx 1.64\times 10^3}$ & ${\approx 1.64\times 10^3}$ & ${\approx 5.71\times 10^3}$ & $\mathbf{\approx 1.64\times 10^3}$\\ 
 & {Computation time (s)}& ${\approx 2.47\times10^4}$ & ${\approx 1}$ & ${\approx 4.97}$ & $\mathbf{\approx 0.19}$\\ \midrule%\hline
 \multirow{3}{*}{\hspace{2.5mm}\rotatebox{0}{\hspace{-0mm}\textbf{Scenario} $\mathbf{2}$}}%&{No. servers ($|\mathcal{R}|+1$)} & $11$ & $11$ & $11$ & $11$ \\
 %& {No. RSUs}& $10$ & $10$ & $10$ & $10$ \\ 
% & {No. requesting vehicles (agents) ($|\mathcal{I}|$)} & $100$ & $100$ & $100$ & $100$ \\ 
% & {Vehicles' speed in lanes \{$1,4$\}, \{$2,5$\}, \{$3,6$\} (km/h) } & $90$, $100$, $120$ & $90$, $100$, $120$ & $90$, $100$, $120$ & $90$, $100$, $120$\\ 
% & {Inter-RSU distance ($D_{\text{R}}$) (km)} & 3 & 3 & 3 & 3 \\
 & Forgetting factor ($\lambda$)& N/A & N/A & N/A & $0.5$ \\
 & {Minimum sum of delay and cost} & {N/A} & ${\approx 1.71\times 10^4}$  & ${\approx 5.3\times 10^4}$ & $\mathbf{\approx 1.71\times 10^4}$\\ 
 & {Computation time (s)}& {N/A} & ${\approx 147.78}$ & ${\approx 145.8}$ & $\mathbf{\approx 68.58}$\\ \bottomrule %\hline
\end{tabular}
\end{center}
\label{table:comparison}
%\vspace{-0.5cm}
\end{table*}

To this end, we will now introduce a forgetting factor for updating $\bar{D}_i^{(t)} (j,k)$ as:
\begin{equation}
\label{eq:regretmatching_recursive_forgetting}
\bar{D}_i^{(t)} (j,k) = \lambda \ \bar{D}_i^{(t-1)} (j,k) + (1-\lambda)\ D_i^{(t)} (j,k),
\end{equation}
where $0\leq \lambda \leq 1$ is a forgetting factor used to regulate the influence of outdated values of regret over the instantaneous regret.
%\begin{figure}[t]
%        \centering
%        \includegraphics[width=0.4\textwidth]{simulation_results/Result1.eps}
%        \captionsetup{justification=centering}
%        \caption{Convergence performance of the four schemes.}
%        \label{fig:result1}
%\end{figure}
Each player then independently chooses its next action according to the following probabilities:\footnote{$|x|^+ = \max\{x,0\}$ for a real number $x$.}
\begin{align}
\pi_i^{(t+1)}(k) & = \frac{1}{\mu} \ | \bar{D}_i^{(t)} (j,k) |^+ \ ,
\label{eq:CRM}
\end{align}
for all $k \neq j$, and $\mu$ is chosen such that the probability of playing the same action in the next iteration is positive. The pseudo-code of our proposed distributed algorithm, which runs independently by each agent, is shown in Algorithm~\ref{alg:RMalg}. Our main theorem is as follows.%\footnote{Reviewers can find a sketch of the proof and the code in the appendix.}

\begin{theorem}
If {\it every} player chooses their actions according to Algorithm~\ref{alg:RMalg}, then the joint empirical distribution of action profiles converges almost surely to the \textit{Correlated Equilibrium} set of the game $\mathcal{G}$ as $t \rightarrow \infty$.\footnote{The proof is given in the technical appendix.}
\end{theorem}
%\begin{proof}
%    The proof is given in the technical appendix. %due to space constraints.
%\end{proof}

%========================================================================================================================
%The proof follows similar steps to those in~\cite{Hart_2001}  and for space constraint, we skip the details here. As also discussed in~\cite{Hart_2001}, if convergence to a point does occur then that point is a pure NE.

%\ed{HN: we should also say something about sub-linear or square root convergence speed here. The proofs for these results probably will make it into a journal paper later.}

%\footnote{Please increase the size of the figures. We can fill the whole 6 pages and submit as a correspondence.}

\section{Performance Evaluation} \label{sec:simulation results}

%\begin{figure*}[t]
%\centering
   %\vspace{-0.5cm}

\subsection{Simulation Settings}

We evaluate the performance of our proposed Algorithm~\ref{alg:RMalg} through numerical experiments in MATLAB (ver. 2021b). The experiments are implemented on a PC with an AMD Ryzen 9 5900X@3.7GHz (24 CPUs) core and 32GB of RAM. We consider a six-lane highway with three lanes in each direction, as depicted in Figs.~\ref{fig:server_1} and~\ref{fig:server_2}. Similar to \cite{QYuan2020,LTTan2018,ZNing2020}, we set $|\mathcal{R}|\in\{2,10\}$, $|\mathcal{I}|\in[25;100]$, $D_{\text{R}} \in\{3,6,9\}$~km, $R_{\text{R}}\in[500;600]$~m, $s_i=200$~MB, $\theta=500$ MB, $f_i\in[0.5;1.2]$~Gcycles, $B_{\text{R}}=100$~MHz, $\alpha=0.02$~s/hop, $\delta^{r,\hat{r}}_{\text{m}}=0.002$~unit/MB, $p_i=20$~dBm and $F_{r,i}\in[1;3]$~GHz. If $r\in\{\text{BS}\}$, then we set $B_{r,i}=0.25$~MHz, $\delta^r_{\text{u}}=20$~units/MHz, $F^{\max}_r=30$~GHz and $\delta^r_{\text{p}}=100$~units/GHz. Otherwise, we set $B_{r,i}=1$~MHz, $\delta^r_{\text{u}}=2$~units/MHz, $F^{\max}_r=20$~GHz and $\delta^r_{\text{p}}=10$~units/GHz. We use $\beta=\gamma=1$ and $\lambda\in\{0.5,0.99,0.9999\}$. In addition, the time step $t$ is set as $1$ s. Similar to \cite{NSW2019} and \cite{BLNguyen2020P}, the vehicle speeds in lanes $\{1, 4\}$, $\{2, 5\}$ and $\{3, 6\}$ are set as $40$ or $90$, $50$ or $100$, and $60$ or $120$ km/h, respectively. 

\begin{figure}[t]
%\centering
%    \begin{minipage}[!h]{0.4\textwidth}
    \centering
        \includegraphics[width=0.35\textwidth]{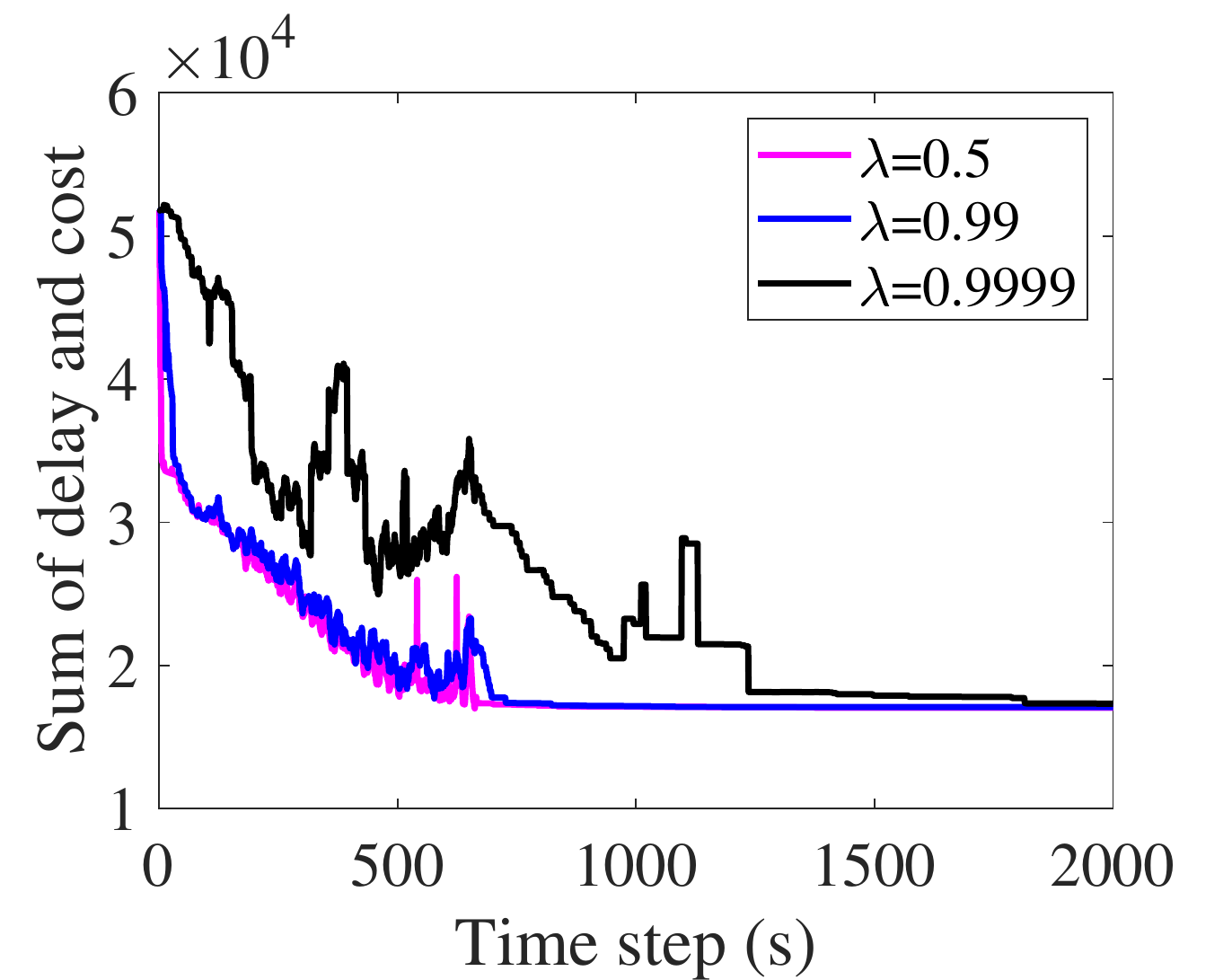}
        \caption{Convergence performance of the proposed scheme when varying the values of forgetting factor $\lambda$ ($D_{\text{R}}=3$ km, $11$ servers, $100$ requesting vehicles (agents), and vehicles' speed of $90$, $100$ and $120$ km/h in lanes \{$1,4$\}, \{$2,5$\} and \{$3,6$\}, respectively).}
        \label{fig:result2}
%    \end{minipage}
\end{figure}

\begin{figure}[t]
%    \begin{minipage}[!h]{0.4\textwidth}
    \centering
        \includegraphics[width=0.35\textwidth]{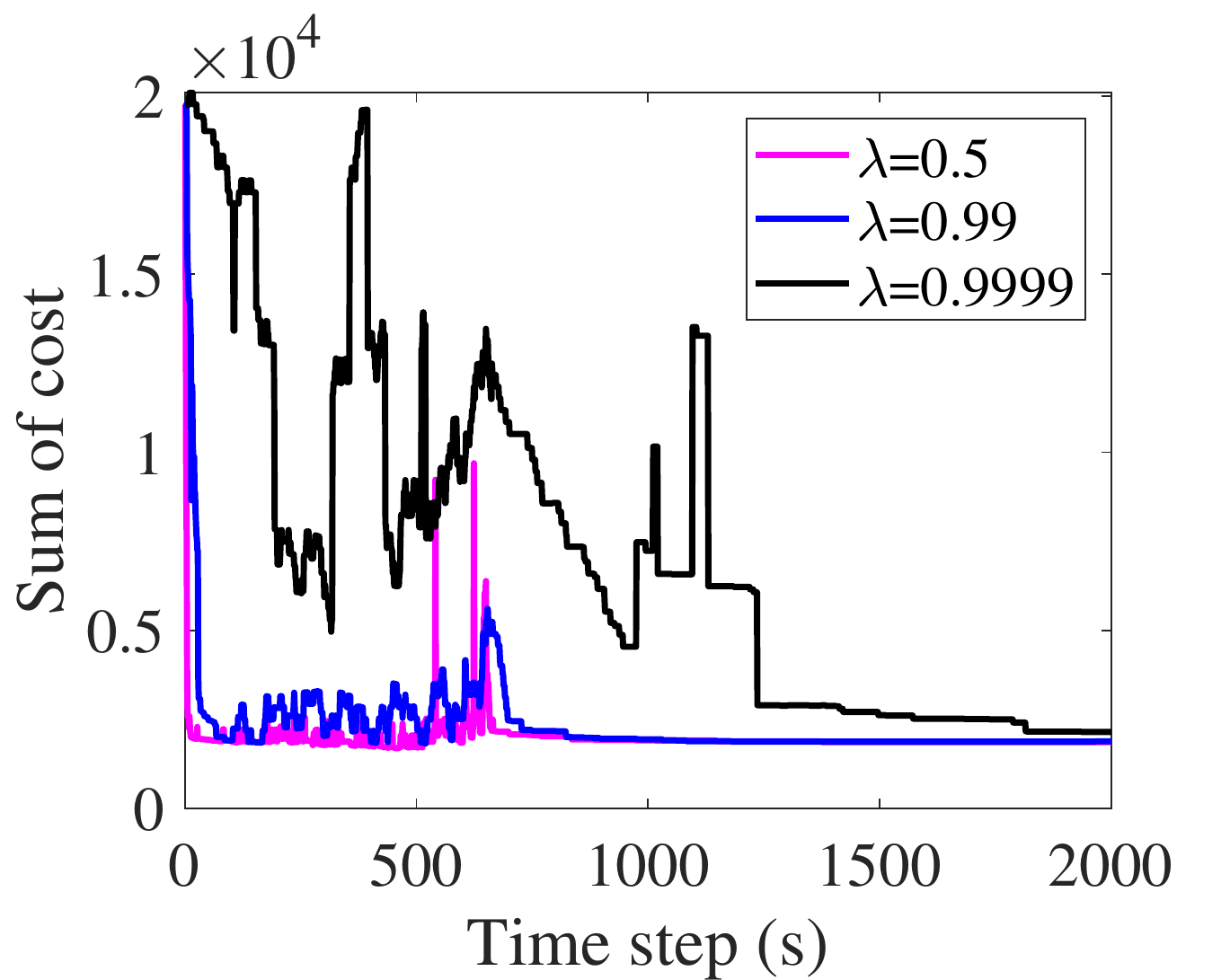}
        \caption{{Total cost of the proposed scheme when varying the values of forgetting factor $\lambda$ ($D_{\text{R}}=3$ km, $11$ servers, $100$ requesting vehicles (agents), and vehicles' speed of $90$, $100$ and $120$ km/h in lanes \{$1,4$\}, \{$2,5$\} and \{$3,6$\}, respectively).}}
        \label{fig:result12}
%    \end{minipage}
\end{figure}

\begin{figure}[t]
%    \begin{minipage}[!h]{0.4\textwidth}
    \centering
        \includegraphics[width=0.35\textwidth]{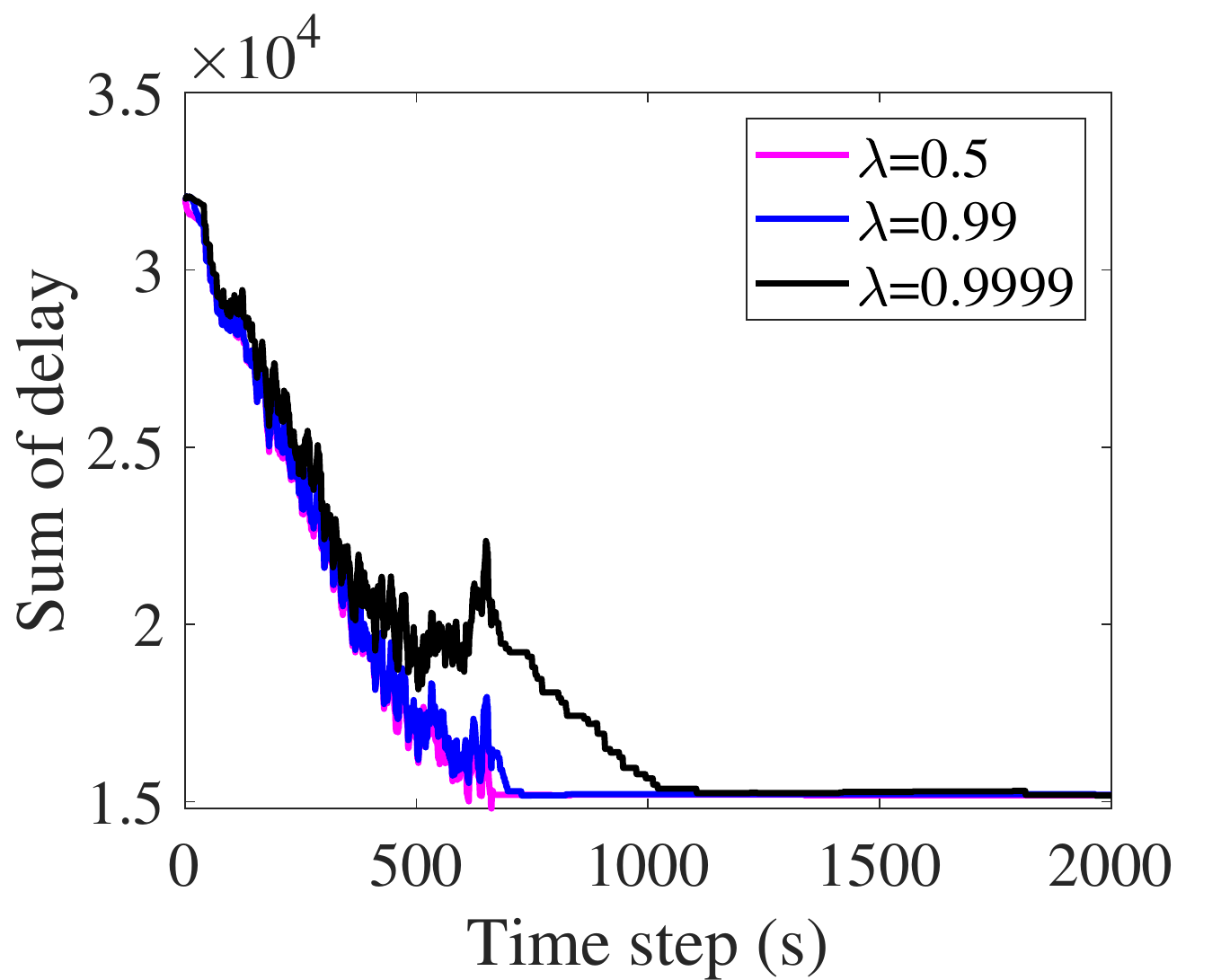}
        \caption{{Total delay of the proposed scheme when varying the values of forgetting factor $\lambda$ ($D_{\text{R}}=3$ km, $11$ servers, $100$ requesting vehicles (agents), and vehicles' speed of $90$, $100$ and $120$ km/h in lanes \{$1,4$\}, \{$2,5$\} and \{$3,6$\}, respectively).}}
        \label{fig:result11}
%    \end{minipage}
\end{figure}

For a comprehensive comparison, we have to benchmark Algorithm~\ref{alg:RMalg} against the most relevant related works in the literature. Different from \cite{HWang2020} and \cite{QYuan2020}, those works must design task assignment algorithms for multi-agent environments while allocating autonomous vehicles' tasks to servers according to their movement in highway scenarios. As a result, we compare Algorithm~\ref{alg:RMalg} with the following three algorithms.

\begin{enumerate}
    \item \textbf{Exhaustive Search (ES)}: A centralized algorithm where a central operator collects all network information and finds the globally optimal solution using exhaustive search, similar to \cite{HWang2020}.
    
    \item \textbf{Traditional Regret-Matching (TRM) scheme} \cite{CFan2020}: A distributed algorithm where {each player selects an action on the basis of its regret value, and the regret values are not calculated with respect to changes in the vehicular network.}  
    
    %\footnote{Duy: a player does not update their own regret function, or it does not send its own updated regret function to other players?} according to changes in the vehicular network.
    
    \item \textbf{Reinforcement Learning with Network-Assisted Feedback (RLNF) scheme} \cite{DDNguyen2017}: A distributed algorithm where each player selects its action without knowing global network conditions.
    
    %an update\footnote{Duy: update of which parameters, from whom and to whom?} on global network conditions.
    
\end{enumerate}
To demonstrate that Algorithm~\ref{alg:RMalg} works effectively in not only small-scale but also large-scale environments, we evaluate all the four algorithms in two scenarios. Specifically, in Scenario~$1$, there are $3$ servers deployed along the highway while $10$ vehicles request to complete their tasks. In contrast to Scenario~$1$, we increase the number of servers and requesting vehicles to $11$ and $100$, respectively, in Scenario~$2$. On the other hand, in both scenarios, the requesting vehicles in lanes $\{1,4\}$, $\{2,5\}$ and $\{3,6\}$ are moving at a speed of $90$, $100$ and $120$ km/h, respectively. In addition, the inter-RSU distance is set as $3$ km.

%\textit{(1)} Exhaustive Search(WM), \textit{(2)} Reinforcement Learning with Network-Assisted Feedback scheme (RLNF) \cite{DDNguyen2017}, and \textit{(3)} Traditional Regret-Matching scheme (TRM) \cite{hart2000simple}.

\subsection{Simulation Results}

Table~\ref{table:comparison} compares the performance of Algorithm~\ref{alg:RMalg} with the three benchmark schemes. {According to the objective function in~\eqref{main problem}, Algorithm~\ref{alg:RMalg} aims to minimize the total delay and cost for task processing; thus, we select the sum of delay and cost as a performance metric.} Here, the total delay plus cost and the computation time for task completion are shown for two scenarios, {as shown in Figs.~\ref{fig:scene_1} and~\ref{fig:scene_2}}. In Scenario~$1$, by using $3$ servers, Algorithm~\ref{alg:RMalg}, ES and TRM complete $10$ tasks with the \emph{lowest} total delay plus cost of $1.64\times 10^3$. 
%\footnote{Duy: what is the unit of this performance metric? And why do we combine delay and cost, but not consider them separately? A justification is needed}. 
Importantly, Algorithm~\ref{alg:RMalg} finds {the best solution} with the \emph{smallest} computation time of $0.19$ s. This represents a significant reduction of more than $80$\% in computation time compared to the next best scheme TRM. Hereby, Algorithm~\ref{alg:RMalg} can be applicable to the real deployment scenarios as it satisfies the latency requirement in vehicular applications, i.e. from $0.1$ s to $0.5$ s \cite{Salman2019,Lin2018}. 

\begin{figure}[t] 
    %\begin{minipage}[t]{0.4\textwidth}
    \centering
        \includegraphics[width=0.35\textwidth]{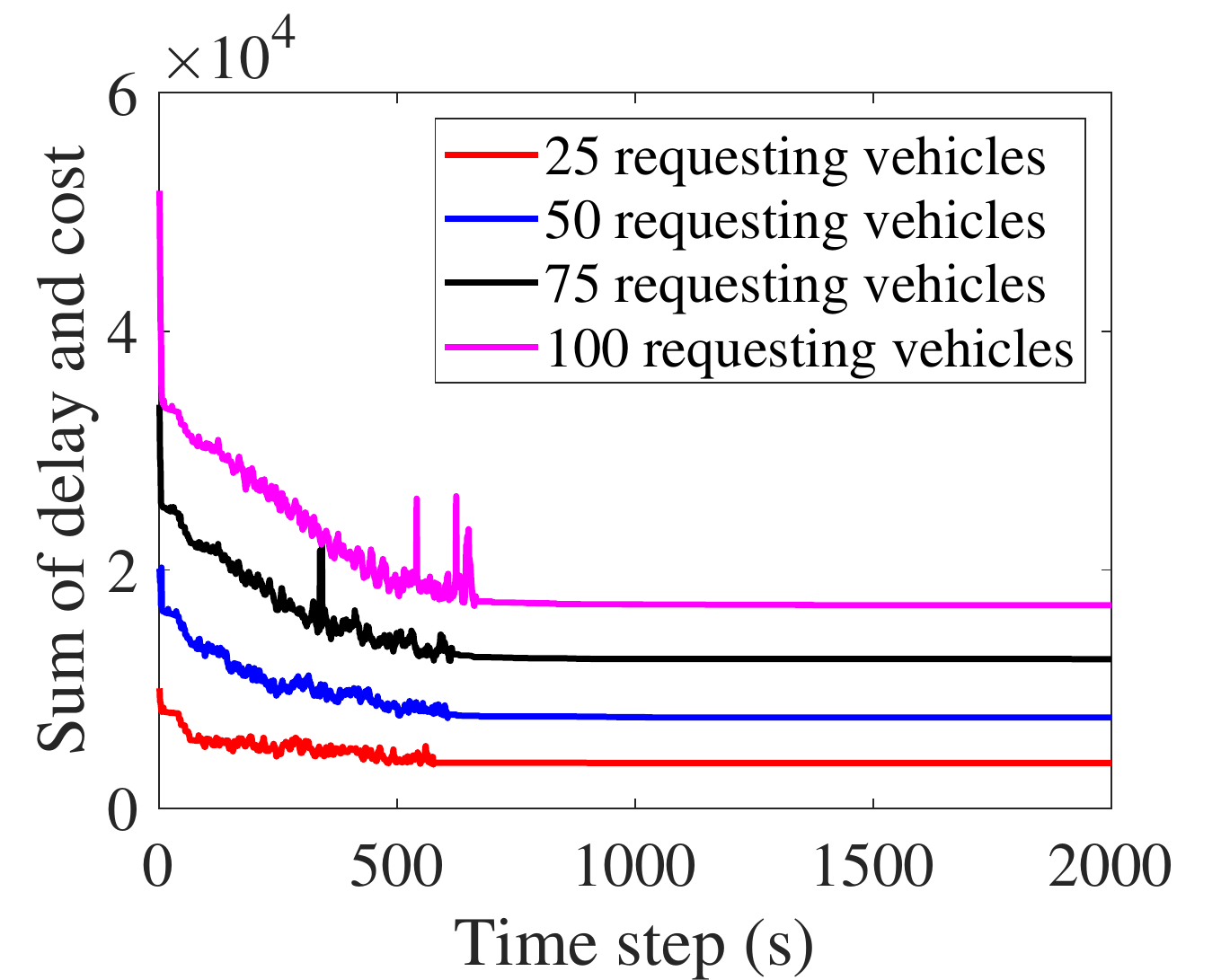}
        \caption{Convergence performance of the proposed scheme when varying the number of requesting vehicles ($\lambda=0.5$, $D_{\text{R}}=3$ km, $11$ servers, and requesting vehicles' speed of $90$, $100$ and $120$ km/h in lanes \{$1$,$4$\}, \{$2$,$5$\} and \{$3$,$6$\}, respectively).}
        \label{fig:result4}
    %\end{minipage}
\end{figure}

\begin{figure}[t]
   %\begin{minipage}[t]{0.4\textwidth}
   \centering
        \includegraphics[width=0.35\textwidth]{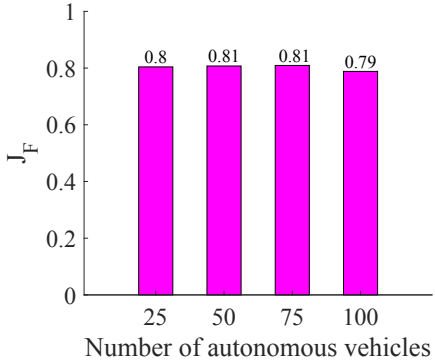}
        \caption{Utility fairness among agents (autonomous vehicles)  when varying the number of autonomous vehicles ($\lambda=0.5$, $D_{\text{R}}=3$ km, $11$ servers, and requesting vehicles' speed of $90$, $100$ and $120$ km/h in lanes \{$1$,$4$\}, \{$2$,$5$\} and \{$3$,$6$\}, respectively).}
        \label{fig:fairness}
    %\end{minipage}
   %\vspace{-0.3cm}
\end{figure}

\begin{figure}[t]     
    %\begin{minipage}[t]{0.4\textwidth}
    \centering
        \includegraphics[width=0.35\textwidth]{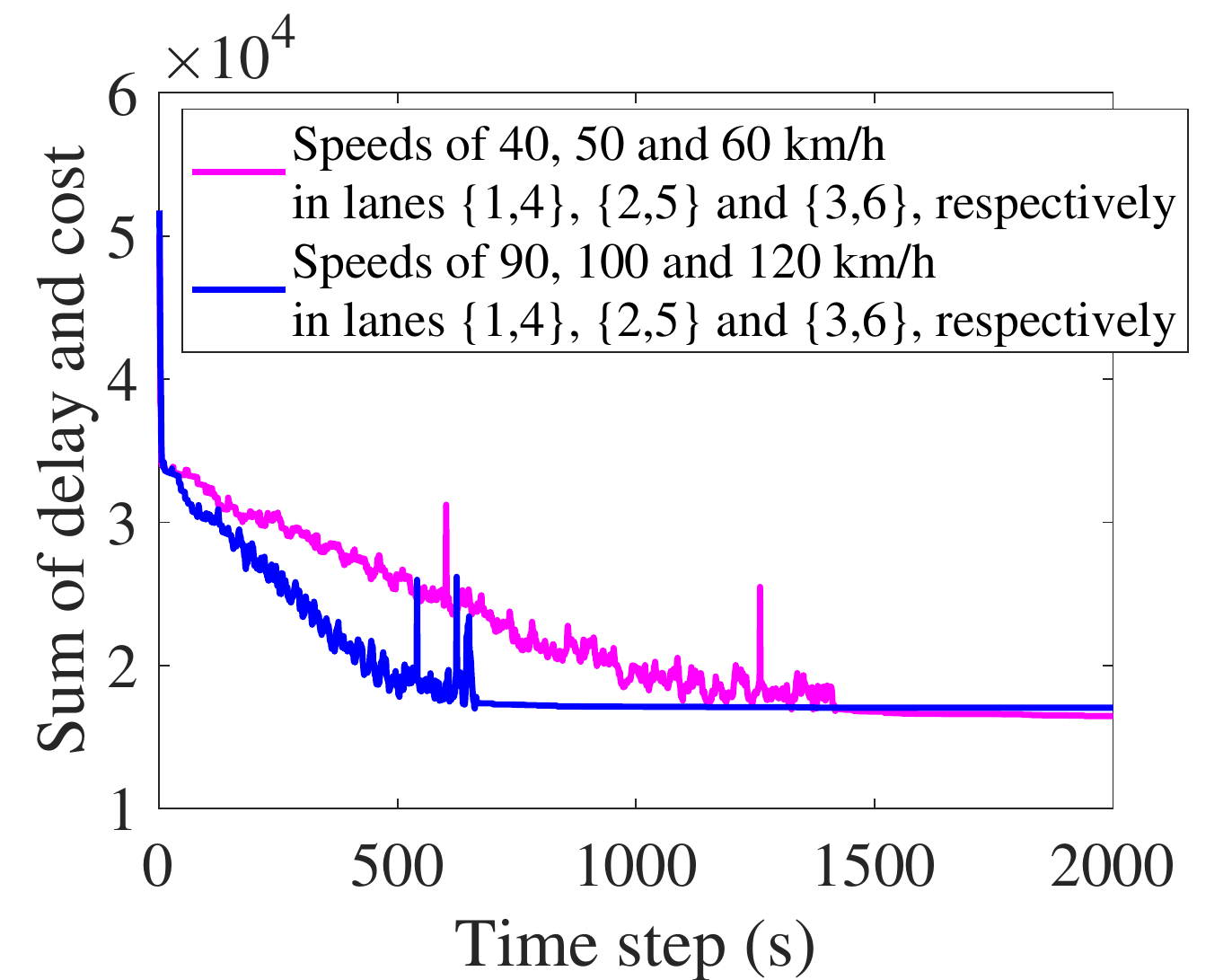}
        \caption{Convergence performance of the proposed scheme when varying the speeds of requesting vehicles ($11$ servers, $100$ requesting vehicles (agents), $\lambda=0.5$ and $D_{\text{R}}=3$ km).}
        \label{fig:result3}
    %\end{minipage}
\end{figure}

\begin{figure}[t]
   %\begin{minipage}[t]{0.4\textwidth}
   \centering
        \includegraphics[width=0.35\textwidth]{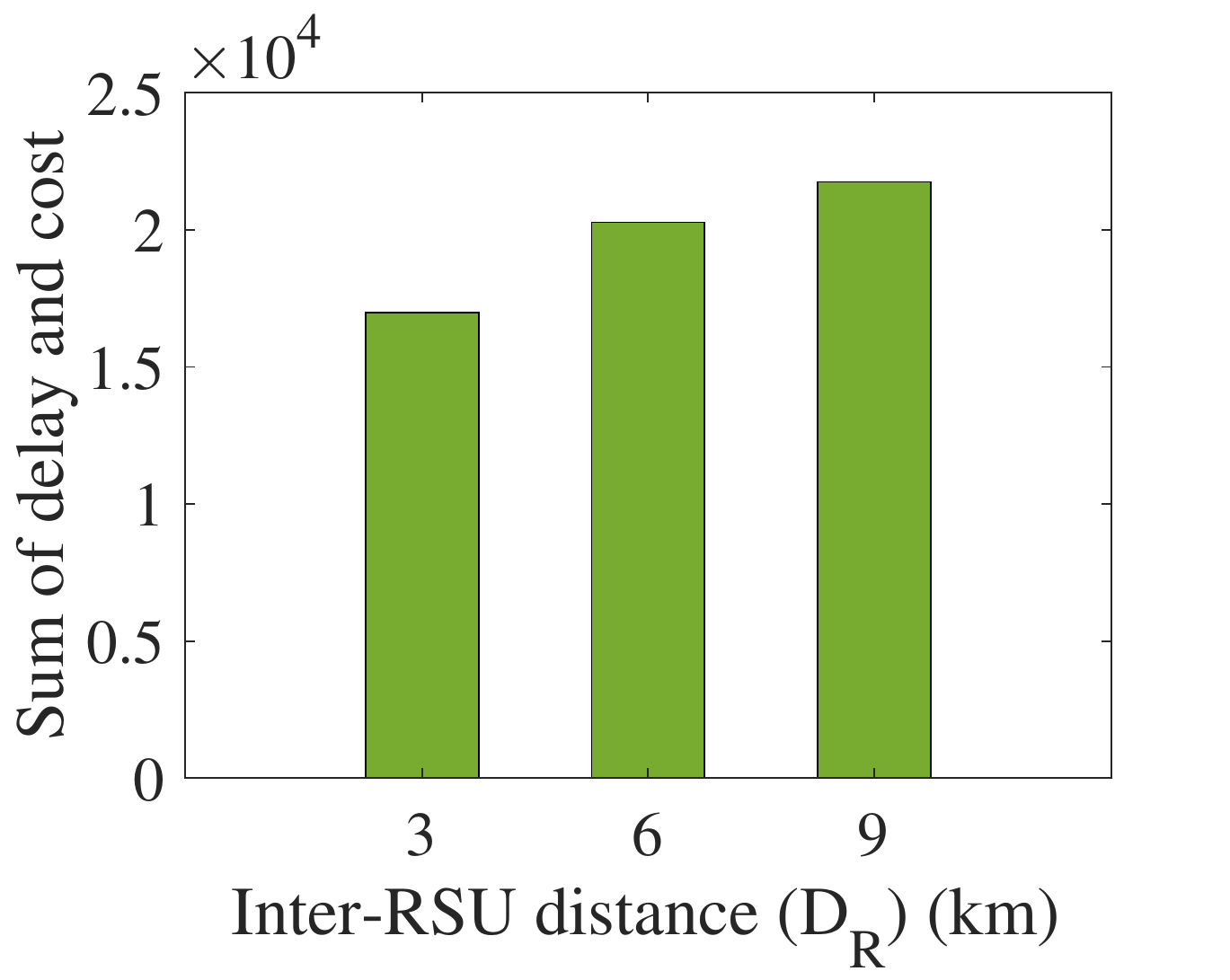}
        \caption{{Performance of the proposed scheme when varying the inter-RSU distance $D_{\text{R}}$ ($\lambda=0.5$, $11$ servers, $100$ requesting vehicles (agents), and vehicles' speed of $90$, $100$ and $120$ km/h in lanes \{$1,4$\}, \{$2,5$\} and \{$3,6$\}, respectively).}}
        \label{fig:result5}
    %\end{minipage}
   %\vspace{-0.3cm}
\end{figure}

In Scenario~$2$, the number of servers and tasks is increased up to $11$
%\footnote{Duy: ``up to" or ``to"? I do not see ``11" anywhere in the table.} 
and $100$, respectively. Given the specifications of a typical PC, it is impossible for ES to iterate through $(11^{100} \approx 1.38\times 10^{104})$ potential solutions. Of the remaining three algorithms, Algorithm~\ref{alg:RMalg} converges to {the CE solution} within the shortest computation time of $68.58$ s, giving the lowest total delay plus cost of $1.71\times 10^4$. {In general, a CE solution might not necessarily be the optimal solution for the BNLP problem~\eqref{main problem}. Here, we show through experiment results  that in most realistic networks, the gap between CE and optimal solution is small (almost negligible) --- illustrating that Algorithm 1 provides a good trade off between convergence speed and optimality.} It is noted that the TRM and RLNF schemes are not able to perform as well, despite they are also based on RM learning. The reason is that the regret values in the TRM and RLNF schemes are not updated with respect to changes in the operating vehicular environment. Furthermore, since the TRM scheme makes task assignment decisions based on information about global network conditions, its total delay plus cost is lower than that of RLNF. 

%\footnote{[HN] we should make a comment here that: a CE solution might not necessarily correspond to the optimal solution for the BNLP problem. Experiment results however show that in most realistic networks, the gap between CE and optimal solution is small (negligible) illustrating that Algorithm 1 is a good trade off between speed and optimality.}. 

%Using our PC with a normal configuration, 

%It is impossible to 

%Fig.~\ref{fig:result1} depicts the total delay and cost of completing $50$ tasks. 

%In contrast to WM, the tasks in RLNF, TRM and Algorithm~\ref{alg:RMalg} are migrated across the RSUs/BS.

%It is clear that Algorithm~\ref{alg:RMalg} converges to the best solution here---the CE solution.  

%In the WM scheme, the constraints of computing capacity and completion delay in problem~\eqref{main problem} are violated because task migration does not happen. As seen from Fig.~\ref{fig:result1}, WM is the worst performer.

The fast convergence behaviour of the proposed scheme is essential for a dynamic environment in which ITS applications operate. {Figs.~\ref{fig:result2},~\ref{fig:result11} and~\ref{fig:result12} illustrate how the convergence of Algorithm~\ref{alg:RMalg} depends on the forgetting factor $\lambda$. In addition, Figs.~\ref{fig:result11} and~\ref{fig:result12} show the impact of $\lambda$ on delay and cost, respectively.} As seen, the fastest convergence occurs when $\lambda$ decreases from $0.9999$ to $0.5$. At $\lambda=0.5$, the cumulative regret of an agent is updated with respect to their instantaneous regret rather than their outdated regret. 
%We choose $\lambda=0.5$ in the implementation of Algorithm~\ref{alg:RMalg} in this paper.
Furthermore, Fig.~\ref{fig:result4} shows that such fast convergence is always maintained irrespective of the number of participating agents. 

To quantify fairness in terms of utility among agents (autonomous vehicles), we make use of Jain's fairness index proposed in \cite{Jain1998AQM} as follows:
\begin{equation}
    J_{\text{F}}=\frac{\left[\displaystyle\sum_{i \in \mathcal{I}} u_i^{(t)}\left(a_i^{(t)},a_{-i}^{(t)}\right)\right]^2}{|\mathcal{I}|\displaystyle\sum_{i \in \mathcal{I}} {u_i^{(t)}\left(a_i^{(t)},a_{-i}^{(t)}\right)}^2 }.
\end{equation}
In addition, the utility fairness among the agents would be maintained when the approximate value of $J_{\text{F}}$ is $1$. As shown in Fig.~\ref{fig:fairness}, Algorithm~\ref{alg:RMalg} achieves the fairness $J_{\text{F}}$ close to $1$, e.g. $0.79$, even though the autonomous vehicle density is high. Here, to calculate $J_{\text{F}}$, we employ the value of agents' utility when Algorithm~\ref{alg:RMalg} converges at the correlated equilibrium.

Fig.~\ref{fig:result3} demonstrates that Algorithm~\ref{alg:RMalg} adapts quickly to the environment changes caused by the agents' mobility. Here, when the agents move at high speeds, it causes a decrease in the duration when they pass through an RSU, or they will enter the next RSUs' coverage range. Thus, the number of agents' actions which are able to both minimize the sum of delay and cost, and satisfy constraints in~\eqref{main problem} is reduced significantly. Certainly, it would be much less than that of the scenario in which the agents traverse the highway with lower speeds. As a result, the convergence speed of Algorithm~\ref{alg:RMalg} in the former is quicker than that in the latter. In particular, Fig.~\ref{fig:result5} shows the influence of RSU deployment on the performance of our proposed scheme. With the shortest distance between two consecutive RSUs (i.e. $D_{\text{R}}=3$ km), the RSUs are distributed densely along the six-lane highway. It leads to the lowest total delay and cost achieved by Algorithm~\ref{alg:RMalg}. Extending the inter-RSU distance will cause an increase in the total delay and cost.

%It makes the agents' action selection restricted. 

%\footnote{How? It is not clear from the figure how the quick adaptation occurs. Can we explain why the algorithm converges faster for higher vehicle speeds? I would expect the reverse, i.e., faster convergence for slower vehicle speeds.!}. % In particular, an increase in the number of agents makes the total delay and cost of task completion higher. 

%that as the number of participating vehicles (equivalently, the number of tasks to be processed) grows, Algorithm~\ref{alg:RMalg} achieves an even much faster convergence speed than does the second-best TRM algorithm. This important result demonstrates the advantage of our proposal in large-scale problems, which are common in vehicular networks.  

%Fig.~\ref{fig:result2} shows that as the inter-RSU distance decreases, the objective value of problem \eqref{main problem} also decreases. This result confirms that one must increase the density of RSU deployment (i.e., decrease the inter-RSU distance) to reduce total delay and cost of task offloading. In that case, the vehicles would have more chance to completely process their tasks at their serving RSUs, resulting in a shorter delay and a lower cost. In any case, Algorithm~\ref{alg:RMalg} achieves the best performance.%, with a total delay and cost of $10,193$ at $D_{\text{R}}=3$~km.

%%=============================================================================================================
\vspace{-3mm}\section{Conclusion} \label{sec:conclusion}
To address the issue of limited computation resources in VEC, this paper proposes a task assignment scheme where vehicles' tasks are migrated across multiple servers according to their mobility pattern. To this end, we have formulated a BNLP problem that minimizes the total delay and cost incurred by the participating vehicles. We then proposed a multi-agent RM learning-based algorithm that is theoretically proved to converge to the CE solution of the formulated problem. The simulation results show the clear advantages of our proposed algorithm over existing solutions.

% \balance

\appendix

%\section*{Technical Appendix}

% ========================================================================================================================
\section*{Proof of Theorem 1}
\begin{proof}
	For notational convenience, let us drop the subscript $i$ and define the following Lyapunov function:\footnote{$\mbox{dist}(x,\Ac) = \min\{\|x-a\|: a\in \Ac\}$, where $\|\cdot\|$ is the Euclidean norm.}
	\begin{equation}
		\label{eq:Lyapunov}
		P(\bar{D}) = \frac{1}{2}\left(\mbox{dist}[\bar{D},\Rbb^-]\right)^2 = \frac{1}{2}\sum\nolimits_{j,k}\left(|\bar{D}(j,k)|^+\right)^2 \ ,
	\end{equation}
	where $\Rbb^-$ represents the negative orthant. Taking the time-derivative of~\eqref{eq:Lyapunov} yields
	\begin{equation}
		\frac{d}{dt}P(\bar{D})=\sum\nolimits_{j,k}|\bar{D}(j,k)|^+ \times \frac{d}{dt}\bar{D}(j,k)\ .
		\label{eq:Pderivative}
	\end{equation}
	
	First, we find $d\bar{D}(j,k)/dt$ by rewriting~\eqref{eq:regretmatching_recursive_forgetting} as:
    \begin{align}
        & \bar{D}^{(t)}(j,k) \nonumber\\ 
        & = \bar{D}^{(t-1)}(j,k)+(1-\lambda)\Big\{D^{(t)}(j,k)-\bar{D}^{(t-1)}(j,k)\Big\}\nonumber\\
        & = \bar{D}^{(t-1)}(j,k)+(1-\lambda) \Big\{\big[u(k,\cdot)-u(j,\cdot)\big] \mathbbm{I}{\{a_i^{(t)} = j\}} \nonumber\\
        & \quad  -\bar{D}^{(t-1)}(j,k)\Big\} \ .
    \label{eq:Rrecursive}
    \end{align}
    Let $\epsilon = 1-\lambda$ be a constant step size. It can be seen that~\eqref{eq:Rrecursive} has the form of a constant step size stochastic approximation algorithm $\theta_{k+1}=\theta_k+\epsilon H(\theta_k,x_k)$ and satisfies \cite[Th.~17.1.1]{krishnamurthy2016partially}. Thus, its dynamics can be characterised by an ordinary differential equation (see \cite[Ch. 17]{krishnamurthy2016partially} for further details). This means the system can be approximated by replacing $x_k$ with its expected value. By applying \cite[Theorem 17.1.1]{krishnamurthy2016partially}, $\bar{R}_t(j,k)$ converges weakly (in distribution) to the averaged system corresponding to~\eqref{eq:Rrecursive}. As such,
    \begin{align}
        \frac{d}{dt}\bar{D}(j,k) 
        & = \Er \Big\{ \big[ u(k,\cdot)-u(j,\cdot)\big] \mathbbm{I}{\{a_i^{(t)} = j\}} - \bar{D}(j,k) \Big\} \nonumber\\
        & = \big[u(k,\cdot) - u(j,\cdot)\big]\ \pi(j)\ - \bar{D}(j,k) \ .
        \label{eq:Rderivative}
    \end{align}

    Next, replacing $d\bar{D}(j,k)/dt$ from~\eqref{eq:Rderivative} into~\eqref{eq:Pderivative} gives:
    \begin{align}
        \frac{d}{dt}P(\bar{D}) &= \sum\nolimits_{j,k}|\bar{D}(j,k)|^+ \Big[u(k,\ell)-u(j,\ell)\Big]\ \pi(j)\nonumber\\
        & \quad - \sum\nolimits_{j,k}|\bar{D}(j,k)|^+ \times \bar{D}(j,k) \nonumber\\
        & \leq \frac{2G\delta}{|\Ac_i|} \sum\nolimits_{j,k}|\bar{D}(j,k)|^+ - 2P(\bar{D}) \ ,
        \label{eq:Pderivative2}
    \end{align}
    where $G$ is an upper bound on $|u(\cdot)|$, $0\leq \delta \leq 1$, and $|\Ac_i|$ is the cardinality of the set $\Ac_i$ (the set of actions of a player $i$). Note that in the last equality of~\eqref{eq:Pderivative2}, we have used the following two lemmas:\footnote{The proof of Lemma~2 is similar to the proof of Theorem 5.1 in~\cite{DDNguyen2018}, so the proof is omitted here.}
    $$(1)\ \sum_{j,k}|\bar{D}(j,k)|^+ \bar{D}(j,k)=2P(\bar{D})\ \mbox{(immediate from Eq.~\eqref{eq:Lyapunov})}$$
    $$(2)\ \sum_{j,k}|\bar{D}(j,k)|^+ \big[u(k,\cdot)-u(j,\cdot)\big]\ \pi(j) \leq \frac{2G\delta}{|\Ac_i|} \sum_{j,k}|\bar{D}(j,k)|^+ .$$
    
    Finally, it follows from~\eqref{eq:Pderivative2} that by assuming $|\bar{D}(j,k)|^+\geq \kappa > 0$, one can choose a sufficiently small $\delta >0$ such that $$\frac{d}{dt}P(\bar{D}) \leq - P(\bar{D})\ .$$
    This implies that $P\big(\bar{D}_i^{(t)}\big)$ goes to zero at an exponential rate. Therefore, 
	$\displaystyle\lim_{t\rightarrow\infty}\mbox{dist} \big[\bar{D},\Rbb^-\big]=0$. %, i.e., convergence to the negative orthant.
	
	Let $\phi_t$ be the empirical distribution of the joint action $(j,a_{-i}^{(t)})$ by all players up to the time $t$. It can be defined by a stochastic approximation recursion as: %, where $j$ is the action of player $i$ and $a_{-i}^{(t)}$ is the actions of the others at time $t$
	\begin{align}
		\phi_t\lb j,a_{-i}^{(t)}\rb & =\phi_{t-1}\lb j,a_{-i}^{(t-1)}\rb \nonumber\\ 
		&\quad + \epsilon\lsb \mathbbm{I}{\lbr a^{(t)} = \lb j,a_{-i}^{(t)}\rb \rbr} - \phi_{t-1}\lb j,a_{-i}^{(t-1)} \rb \rsb \nonumber\\
		&=\epsilon \sum_{\tau\leq t}  (1-\epsilon)^{t-\tau}\ \mathbbm{I}{\lbr a^{(t)} = \lb j,a_{-i}^{(t)}\rb \rbr}\ .
		\label{eq:globalbehaviour}
	\end{align}
	The elements of the regret matrix in~\eqref{eq:regretmatching_recursive} can be rewritten as follows %The result of Theorem 1 is immediate from the definition of the ``regret". using the non-recursive expression
	\begin{align}\label{eq:D_final}
		& \bar{D}_i^{(t)}(j,k) 
		= \epsilon \sum_{\tau\leq t} (1-\epsilon)^{t-\tau} \Big[ u_i(k,\cdot)-u_i(j,\cdot) \Big] \mathbbm{I}\lbr a_i^{(\tau)}=j\rbr\nonumber\\
		&= \sum_{a_{-i}} \epsilon \sum_{\tau\leq t} (1-\epsilon)^{t-\tau} \ \mathbbm{I}\lbr a_i^{(\tau)}=j\rbr \ \pi_{-i}^{(t)} \Big[ u_i(k,\cdot)-u_i(j,\cdot)\Big]\nonumber \\
		&= \sum_{a_{-i}} \epsilon \sum_{\tau\leq t}(1-\epsilon)^{t-\tau} \ \mathbbm{I}{\lbr a^{(t)} = \lb j,a_{-i}^{(t)}\rb \rbr} \Big[ u_i(k,\cdot)-u_i(j,\cdot)\Big]\nonumber\\
		&= \sum_{a_{-i}}\phi_t\lb j,a_{-i}^{(t)}\rb \Big[ u_i(k,\cdot)-u_i(j,\cdot)\Big]\ .
	\end{align}
	On the last line of \eqref{eq:D_final}, we have substituted $\phi_t\lb j,a_{-i}^{(t)})\rb$ from~\eqref{eq:globalbehaviour}.
	Finally, on any convergent subsequence $\displaystyle\lim_{t \rightarrow \infty} \phi_t \rightarrow \psi$, we have: 
	\begin{align}\label{eq:final_proof}
	    \lim_{t \rightarrow \infty} \bar{D}_i^{(t)}(j,k) = \sum_{a_{-i}} \psi(j,a_{-i}) \big[ u_i(k,a_{-i}) - u_i(j,a_{-i}) \big] \leq 0\ .
	 \end{align}
	Comparing~\eqref{eq:final_proof} with the definition of Correlated Equilibrium in Eq.~\eqref{eq:correlatedequilibrium} completes the proof.
\end{proof}

\bibliographystyle{IEEEtran}

\bibliography{reference}

\balance

\end{document}